\documentclass[pre,twocolumn,floatfix,superscriptaddress,aps]{revtex4}
\usepackage{amsmath}
\usepackage{makeidx}
\usepackage{amssymb}
\usepackage{graphicx}

\usepackage[usenames]{color}
\usepackage[normalem]{ulem}

\usepackage{hyperref}
\usepackage[T1]{fontenc} 

\begin{document}

\title{Phase-separation of  vector  solitons in  spin-orbit coupled 
       spin-$1$ condensates}

\author{S. K. Adhikari\footnote{sk.adhikari@unesp.br }}
\affiliation{Instituto de F\'{\i}sica Te\'orica, Universidade Estadual
             Paulista - UNESP, \\ 01.140-070 S\~ao Paulo, S\~ao Paulo, Brazil}
      

\date{\today}
\begin{abstract}
We study the   phase-separation in three-component    bright vector solitons in a 
quasi-one-dimensional spin-orbit-coupled hyper-fine spin $F=1$ ferromagnetic 
Bose-Einstein 
condensate 
upon an increase of the strength of 
spin-orbit (SO) coupling $p_x \Sigma_z$ above a critical value, where $p_x$ is the linear momentum and $\Sigma_z$ 
is the $z$-component of the spin-1 matrix. The bright vector  solitons are demonstrated to be mobile 
and dynamically stable.     The collision between two such vector solitons is found to be elastic at 
 {\it all} velocities 
with the conservation of density of each vector soliton. The two colliding vector 
solitons repel at small separation and at very small colliding velocity,  they come close  and bounce back with the same velocity without ever encountering each other.
This repulsion  produced by SO coupling is responsible for 
the phase separation in a vector soliton for large strengths of SO coupling.   
{ The collision dynamics is found to be completely insensitive to the relative phase of the colliding solitons.} However, in the absence of SO coupling, at very small velocity, the two colliding vector solitons attract 
each other and form a vector soliton molecule and the collision dynamics is sensitive to the relative phase as in scalar solitons.  
The present  investigation is carried out through a numerical solution and an analytic variational approximation 
of the underlying mean-field Gross-Pitaevskii equation.

\end{abstract}

\maketitle

\section{Introduction}
\label{Sec-I}
 Bright solitons are  self-bound solitary wave that can move at a 
constant velocity maintaining  its shape due to a cancellation 
of   linear repulsion   and  non-linear  attraction. Such solitons  have been found \cite{Kivshar}  in  water waves, non-linear optics, and 
Bose-Einstein condensates (BECs)  among others. 
Bright solitons have been created  in a BEC of $^7$Li \cite{li} and $^{85}$Rb \cite{rb} atoms
 by a management of the  non-linear attraction  near a 
Feshbach resonance \cite{Inouye}. 
Solitons have also been studied in  binary 
BEC mixtures  \cite{Perez-Garcia}.  

After the experimental observation of a spinor BEC of $^{23}$Na atoms with hyper-fine spin 
$F=1$ \cite{exp}, mean-field theory to study these have been developed \cite{Ohmi}. 
Although there could not be any natural spin-orbit (SO) coupling in a spinor BEC of {\it neutral} atoms, 
an artificial synthetic   SO coupling can be realized in a spinor BEC by a management of 
external electromagnetic fields  \cite{stringari,rev}. Different managements are possible which lead to a different types of  SO coupling 
between spin and momentum  in the mean-field equation of a spinor BEC.   Two such possible   
SO couplings are due to  Rashba \cite{Rashba} and 
Dresselhaus \cite{Dresselhaus} and other types of SO coupling are possible.  An equal mixture Rashba and Dresselhaus  SO couplings was first realized experimentally in a  pseudo spin-1/2 spinor BEC  formed of   
two ($F_z=0$ and $-1$) of the three  hyper-fine spin components of    
the $F = 1$ state 5S$_{1/2}$ of $^{87}$Rb \cite{Lin}. 
After this pioneering experiment, similar SO-coupled BEC was formed and studied in different 
laboratories \cite{diff}.
Different possible SO couplings in spinor BECs and 
the ways to engineer  these in a laboratory are addressed in review articles   \cite{stringari}.  
Possible ways of realizing the SO coupling in three-component spin-1 BEC have been discussed \cite{SOspin1}. The three components ($F_z =  \pm 1, 0 $) corresponding to  three  spin projections of the spin-1 operator $\Sigma$   will be denoted by the subscripts $\pm 1$ and 0. 

 Solitons have been 
extensively studied in spinor BECs without SO coupling \cite{Ieda}. Novel phases \cite{SOspin1} and 
solitonic structures in SO-coupled pseudo-spin-$1/2$ \cite{rela} and  spin-$1$  
\cite{Liu,sol1d} BECs have also been investigated theoretically. These studies were extended to quasi-solitons confined in two \cite{sol2d}
and three \cite{sol3d} dimensions. 
 Different types of SO coupling introduce rich dynamics through
different types of derivative couplings among the component wave functions of the mean-field model.

A spin-$1$ spinor BEC is controlled  by two interaction strengths, e.g.,
  $c_0\propto (a_0+2a_2)/3$ and $c_2 \propto (a_2-a_0)/3$, with $a_0$ and 
$a_2$ the scattering lengths in total spin $F = 0$ and 
2 channels,  respectively \cite{Ohmi}. All  spin-1 spinor BECs can be classified into  two distinct types \cite{Ohmi,stringari}: 
ferromagnetic    ($c_2<0$) and anti-ferromagnetic ($c_2>0$).  
 In this paper, we study  three-component vector solitons in 
a SO-coupled spin-$1$ ferromagnetic BEC in a
quasi-one-dimensional (quasi-1D) trap  using a mean-field  coupled Gross-Pitaevskii (GP)
equation.

{We consider a distinct SO coupling \cite{usedspin1/2,usedspin1/2x}
  $(\propto \gamma  p_x \Sigma_z)$  where $ p_x$ is the $x$ component of momentum, $\gamma$ the strength of SO coupling, and $\Sigma_z$ is the $z$ component of the spin-1 spin matrix. In a laboratory, this SO coupling can be obtained by two counter-propagating polarized laser fields of slightly different frequencies.
A Raman coupling  due to the lasers
 induces transitions between the three spin components of the spin-1 BEC, providing, at the same time, a momentum transfer along $x$ direction, which determines the strength $\gamma$ of SO coupling.     
As has been shown \cite{usedspin1/2}, 
this SO coupling appears in the single particle Hamiltonian  by applying a unitary transformation to the Hamiltonian in the laboratory frame describing the system in the presence of   detuned, spin-polarized laser fields.
The unitary transformation consists of a local rotation in spin space around the $z$ axis.}
With this SO coupling, we identify novel  
three-component vector solitons in a quasi-1D BEC along the $x$ direction in the  ferromagnetic domain.   For small values of the strength $\gamma$ of SO coupling and also in its absence ($\gamma=0$),  an overlapping 
 three-component vector soliton is formed in a ferromagnetic BEC. With the increase of the strength $\gamma$ of SO coupling above a critical value $\gamma _c$, a phase separation takes place between   the $F_z =\pm 1$ components and the third $F_z= 0$ component vanishes, thus forming a completely phase separated two-component vector  soliton.    Actually, both types of these solitons 
$-$ two-component and three-component $-$ 
exist above and below the critical SO-coupling strength $\gamma=\gamma_c$.  For $\gamma < \gamma_c$, the overlapping soliton is the minimum-energy ground state and the phase-separated soliton is an excited state; whereas, for      $\gamma > \gamma_c$, the opposite is true. 
    In a previous study with a different SO coupling $(\propto p_x \Sigma_x)$ \cite{sol1d} we found distinct types of solitons in the ferromagnetic and anti-ferromagnetic domains.  The ferromagnetic solitons are true mobile overlapping solitons with single-peak structure. The anti-ferromagnetic solitons usually have multi-peak structure and could not move maintaining the shapes of the components.  The present  ferromagnetic  solitons are true mobile solitons.

In the SO-coupled GP equation, we use a plausible analytic approximation and a  variational scheme  to determine the densities of the SO-coupled ferromagnetic 
bright solitons  in  the two above-mentioned domains $-$ overlapping and phase-separated $-$  minimizing the energy functional.  The appropriate 
variational {\em ansatz} in each of the domains is constructed using a knowledge about  the 
solutions of the SO-coupled equation. {The variational analysis
provides  the necessary and sufficient conditions which the interaction strengths $c_0$ and $c_2$
must satisfy to obtain a stable bright soliton.} In addition to the densities and energies of the soliton, the analytic variational method also yields the critical SO-coupling strength  for a phase separation of the components of the vector soliton.     We also compare all these analytic  
variational results with the  numerical solution the 
GP equation obtained by imaginary-time propagation for the stationary vector solitons.
 
 We study the dynamics of the vector soliton numerically by real-time simulation. The dynamical stability of the vector soliton was established.   The phase separation of a three-component vector soliton  was also demonstrated by real-time propagation upon 
changing the SO-coupling strength from a value below $\gamma_c$ to a value above it.  {  We demonstrate that  the present vector soliton can propagate maintaining its shape, although the SO-coupled GP equation is not Galilean invariant.} The collision between two  overlapping SO-coupled ferromagnetic  vector solitons, quite different from three-component solitons without SO coupling, is found to be elastic at  {\it all} velocities  with the conservation of  density. The two vector solitons repel at small distances, and  at small colliding velocities,  they come close and bounce back without  ever encountering each other.  Quite different from two scalar solitons,
the collision dynamics of two SO-coupled vector solitons remain unchanged after a change of relative phase between the two SO-coupled vector solitons. 
To demonstrate that the two unusual properties of collision $-$ (i) elastic nature at all velocities and (ii) insensitivity to relative phase $-$ are caused by the SO coupling, we study the collision dynamics at small velocity of two vector solitons in the absence of SO coupling.  In that case, after collision   the two vector solitons attract and form a vector soliton molecule in an excited state which never separate.  Also, the collision dynamics, with the SO coupling switched off, is found to be very sensitive to the relative phase between the two solitons as in scalar solitons

  In Sec. \ref{Sec-II}, we describe the mean-field model GP equation
for  a SO-coupled spin-$1$ spinor BEC.  We provide analytic variational 
solution of this model for the overlapping vector soliton and for its phase-separated counterpart. In 
Sec. \ref{Sec-III}, we provide a numerical solution of our model for the two 
types of solitons and compare the results for density and energy with the corresponding 
analytic variational results. We also study the dynamics of the vector soliton and 
the transition from an overlapping to phase-separated vector soliton with the increase of 
SO coupling. The collision dynamics of two overlapping vector soliton was also studied at different 
colliding velocities.
  In Sec. \ref{Sec-IV}  a summary of our findings is presented.


\section{Analytical Formulation}
\label{Sec-II}
\subsection{Mean-field model for a SO-coupled BEC}

We consider a SO-coupled spinor BEC in a quasi-1D trap  along the $x$ axis.  
This quasi-1D trap is realized by  strong traps in $y$ and $z$ directions, so that 
the system is frozen in Gaussian ground states in these directions and the essential 
dynamics  of the system is realized in the $x$ direction \cite{Salasnich}.  
The single particle Hamiltonian of the 
condensate  in  this  quasi-1D trap is  taken  in scaled dimensionless units $\hbar={\bar {\bf m}}=1$ as \cite{usedspin1/2} 
\begin{equation}
H_0 = \frac{p_x^2}{2} + \gamma p_x \Sigma_z,
\label{sph} 
\end{equation}
where $\bar {\bf m}$ is the mass of an atom, $p_x = -i\partial/\partial x$ is the momentum operator along $x$
axis,  and $\Sigma_z$ is the irreducible representation of the $z$ component of 
the spin-1 spin matrix:  
\begin{eqnarray}
\Sigma_z=  \left( \begin{array}
 {ccccc}
1 & 0 & 0\\
0 & 0 & 0\\
0 & 0 & -1 \end{array} \right),
\end{eqnarray}
As we will be investigating vector solitons in this paper we will not include any trapping potential 
in the Hamiltonian. 
This SO-coupling is distinct from a previous SO coupling \cite{gautam-1,gautam-2} $(\gamma p_x \Sigma_x)$
used in the study of a quasi-1D 
BEC.

Using the single particle model Hamiltonian  (\ref{sph}) and considering 
interactions in the  Hartree approximation, a quasi-1D \cite{Salasnich} spin-1 BEC 
can be described by the following set of three coupled mean-field partial 
differential GP equations for the wave-function components $\phi_j$ 
\cite{Ohmi}
\begin{align}
 i \frac{\partial \phi_{\pm 1}}{\partial {t}} &=
 \left( -\frac{1}{2}\frac{\partial^2}{\partial  {x}^2}
 +   {c}_0 {\rho}\right)\phi_{\pm 1}\mp {i {\gamma}}
  \frac{\partial\phi_{\pm 1}}{\partial   x}\nonumber\\ 
 &+c_2( {\rho}_{\pm 1}+ {\rho}_0- {\rho}_{\mp 1})\phi_{\pm 1}+ c_2 \phi_0^2\phi_{\mp 1}^*,\label{gps-1}\\
 i\frac{\partial \phi_0}{\partial {t}} &= 
 \left( -\frac{1}{2}\frac{\partial^2}{\partial  {x}^2}
 + {c}_0 {\rho}\right)\phi_0   \nonumber\\
  &  
  +c_2( {\rho}_{+1}+ {\rho}_{-1})\phi_0+ 2 {c}_2\phi_0^*\phi_{+1}\phi_{-1},\label{gps-3}
\end{align}
where interaction strengths \cite{abc2}
$ {c}_0 = 2N (a_0+2a_2)l_0 /3l_{yz}^2 $,
$ {c}_2 = 2N (a_2-a_2) l_0/3l_{yz}^2 $, component density 
$ {\rho}_j = |\phi_j|^2$ with $j=+1,0,-1$ corresponding to the three components of the spin-1 spinor $F_z=+1,0,-1$, 
and total density ${\rho} = (\rho_{+1}+\rho_0+\rho_{-1})$,  $l_{yz}$ is the harmonic oscillator length in the transverse $yz$ directions and $l_0$, in the absense of a trap in $x$ direction, is a scaling length 
in $x$ direction.  The densities are measured in units of $l^{-1}$. 
{ For notational simplicity, in Eqs. (\ref{gps-1})  and (\ref{gps-3}) we have not explicitly shown  the   space and time dependence of the wave function $\phi_{\pm 1,0}(x,t) $. }
The total density is  normalized to unity, i.e., 
$
 \int_{-\infty}^{\infty} {\rho}( {x})d {x} = 1. 
$ The conserved magnetization is defined as $\int 
dx  (\rho_{+1}- \rho_{-1}) =m.$ 
 
  \subsection{Variational Approximation}
\label{II}

The energy functional corresponding to  the mean-field  SO-coupled spinor BEC model 
(\ref{gps-1}) and (\ref{gps-3}) is
\cite{Ohmi}
\begin{align} \label{energy}
 E(\gamma) &=  \int_{-\infty}^{\infty} dx \Bigg\{\frac{1}{2}\left|\frac{d\phi_{+1}}{dx}
  \right|^2+\frac{1}{2}\left|\frac{d\phi_0}{dx}\right|^2+
  \frac{1}{2}\left|\frac{d\phi_{-1}}{dx}\right|^2   +  \frac{c_0}{2}\rho^2  \nonumber\\
  &+  \frac{c_2}{2}\Big[ \left(\rho_{+1} + \rho_0-\rho_{-1} 
  \right)\rho_{+1}+  \left(\rho_0 + \rho_{-1}-\rho_{+1} \right)\rho_{-1}\nonumber\\
   &+ 
  \left(\rho_{+1} + \rho_{-1} \right)\rho_0
+
  2\left( \phi_{-1}^*\phi_0^2\phi_{+1}^*\right. 
  \left.+\phi_{-1}\phi_{0}^{*2}\phi_{+1}\right)  \Big]
\nonumber \\
 &
  + \gamma\left( -i\phi_{+1}^*\frac{d\phi_{+1}}{dx} 
  + i\phi_{-1}^*\frac{d\phi_{-1}}{dx}\right)  \Bigg\}.
\end{align}
 We will minimize this energy functional using an analytic variational wave function 
to find the analytic solution of the SO-coupled GP equation.
 The analytic ansatz for the wave function is taken as 
\begin{equation}
\Phi\equiv  \left( \begin{array}{c} \phi_{+1}\\
\phi_0\\
\phi_{-1}
 \end{array} \right)
= \frac{1}{2}\left( \begin{array}{c}
e^{-i\gamma x}(1+m)\phi(x) \\
\sqrt{2(1-m^2)} \phi(x)\\
e^{i\gamma x}(1-m)\phi(x) \end{array} \right), \quad\\ 
\label{Gaussian}
\end{equation}
where $\phi(x)$ is taken to be a normalized Gaussian
\begin{equation}\label{gauss}
\phi(x) = \left(\frac{1}{\pi \alpha^2}\right)^{1/4}\exp\left[-\frac{x^2}{2\alpha^2}\right],
\end{equation}
 or a hyperbolic secant   function
\begin{equation}\label{sech}
\phi(x) =  \frac{\sqrt \sigma}{\sqrt 2}\mbox{sech} (\sigma x),
\end{equation}
where the parameters $\alpha$ and $\sigma$ denote amplitude and width.  { In the case $\gamma=m=0$, the three components $\phi_{\pm 1}, \phi_0$ are multiples of one another while ansatz (\ref{Gaussian}) becomes  an exact relation and the  hyperbolic secant   function (\ref{sech}) becomes an exact solution of 
Eqs. (\ref{gps-1}) and (\ref{gps-3}). }
With these ansatz for the wave function, the energy functional (\ref{energy}) is explicitly real,  has the correct $\gamma$ dependence, and correct magnetization and normalization.  The same functional form of the wave function components is consistent with the numerical solution of the GP model  and was employed before to predict the component densities of  trapped spin-1 and spin-2 spinor BECs in the form of single mode  \cite{abc} and decoupled-mode \cite{abc2} approximations.  Here we are applying similar ideas to study the properties of a vector soliton.

With the  Gaussian ansatz (\ref{gauss}) for the profile of the vector soliton,  the energy functional
 (\ref{energy}) becomes 
\begin{align}\label{en1}
E(\gamma)=-\frac{\gamma^2}{4}+\frac{1}{4\alpha^2}  +  \frac{c_0+c_2}{2 \alpha \sqrt{2\pi}}.   
\end{align}
This energy functional is independent of magnetization $m$.  The width $\alpha$ of the minimum-energy ground state vector soliton  is obtained by minimizing this energy functional with respect to $\alpha$:  
\begin{align}\label{minal}
\alpha = -\frac{\sqrt{2\pi}}{c_0+c_2}.
\end{align}
For this width to be positive we 
require $c_0+c_2 <0$  in addition to $c_2<0$ (ferromagnetic).   This width  is independent of the SO-coupling  strength  $\gamma$ and also of magnetization $m$.  The following  minimum of energy as
a function of $\gamma$ is  obtained by substituting Eq. (\ref{minal}) in Eq. (\ref{en1})
\begin{align}\label{egy}
E(\gamma) = -\frac{\gamma^2}{4} -  \frac{(c_0+c_2)^2}{8\pi}\approx -\frac{\gamma^2}{4} - 0.0397887(c_0+c_2)^2 ,
\end{align}
which is the energy of the minimum-energy spin-1 three-component vector soliton in the ground state. { For the hyperbolic secant   ansatz (\ref{sech}) for the wave function, energy functional  (\ref{energy}) becomes 
\begin{align}\label{ensi}
E(\gamma)=-\frac{\gamma^2}{4} +\frac{\sigma^2}{6}  +  \frac{(c_0+c_2)\sigma}{6}.   
\end{align}
The minimum of this energy occurs at 
\begin{align}\label{minalsi}
\sigma = -\frac{1}{2}(c_0+c_2),
\end{align}
provided $c_0+c_2 <0.$ The minimum of energy (\ref{ensi}) is 
\begin{align}\label{egysi}
E(\gamma) = -\frac{\gamma^2}{4}-  \frac{(c_0+c_2)^2}{24}\approx -\frac{\gamma^2}{4}- 0.0416667(c_0+c_2)^2.
\end{align}
The energy (\ref{egysi}) obtained with the hyperbolic secant  ansatz (\ref{sech}) is smaller than energy  (\ref{egy}) obtained with the Gaussian ansatz (\ref{gauss}).
Hence, because of the variational nature of the analytic approximation,  the hyperbolic secant   ansatz should give a better approximation, as will be verified in the numerical calculations in Sec. \ref{Sec-III}.
}

 We note that the analytic variational results (\ref{minal}) and (\ref{egy}), {as well as 
(\ref{minalsi}) and (\ref{egysi}), are} determined by the {\it net attraction} $c_0+c_2$ and independent of the
individual interaction strengths $c_0$ and $c_2$.  The numerical results depend in a nontrivial way on the individual strengths $c_0$ and $c_2$, although we will see that in the weak-coupling limit { of small  nonlinear interaction}  the numerical results follow the analytic ones being determined by the net attraction. 
{ The effective nonlinear interaction in the GP equations (\ref{gps-1}) and (\ref{gps-3}) is $\sim |(c_0+c_2)|\rho_{j,\mbox{av}}$, where $ \rho_{j,\mbox{av}}$ is the average density of component $j$.  This condition of weak coupling is valid for the numerical results presented in Sec. \ref{Sec-III}.}

Our numerical calculation revealed that for SO-coupling  strength $\gamma$  larger than  a critical value $\gamma_c$,  
a complete phase separation occurs between the $F_z=\pm 1$ components while the $F_z=0$ component vanishes. 
For $\gamma< \gamma_c$ the   fully-overlapping three-component vector  soliton  is the lowest-energy soliton  and for 
$\gamma> \gamma_c$  the  fully-separated two-component vector soliton is the ground state.  We could not find a partially separated 
vector soliton with non-zero $F_z=0$ component for any values of the parameters: $c_0, c_2$ and $\gamma$.  To study this crossover from three- to two-component ground state soliton 
analytically  we note that if we set $\phi_0=\rho_0=0$,
for vanishing $F_z=0$ component, 
 and the overlap $\rho_{+1}\rho_{-1}=0$, for a complete phase separation,
in Eqs. (\ref{gps-1}) and (\ref{gps-3}), then we get 
the following set of {
decoupled equations for the phase-separated  two-component vector soliton 
\begin{align}\label{gps1} 
 i \frac{\partial \phi_{\pm 1}}{\partial  {t}} =
 \left[ -\frac{1}{2}\frac{\partial^2}{\partial  {x}^2}
+ \kappa_{\pm 1} {(c_0+c_2)}  \rho_{\pm 1}\mp {i {\gamma}}
  \frac{\partial}{\partial   x}   
   \right] \phi_{\pm 1} ,
\end{align}
where $\rho_{\pm 1}=|\phi_{\pm 1}|^2$, $\kappa_{\pm 1}=(1\pm m)/2$. As the coupling between the two components 
has been removed, the nonlinearities $\kappa_{\pm 1}(c_0+c_2)$ are   appropriate 
for magnetization $m$.
Equation (\ref{gps1}) has the following analytic solution  
\begin{align}\label{anali}
\phi_{\pm 1}\equiv\sqrt{ { \kappa_{\pm 1}}} \widehat \phi_{\pm 1}
 =  \sqrt{ { \kappa_{\pm 1}}}  e^{\mp i\gamma x}\sqrt{ \frac{ \sigma_{\pm 1}}{2}}
 \mathrm{sech}(\sigma_{\pm 1}x), 
\end{align}
satisfying the condition of normalization and magnetization, e.g. $\int dx (\rho_{+1}
+\rho_{-1})=1$ and  $\int dx (\rho_{+1}
-\rho_{-1})=m$, respectively, where $\sigma_{\pm 1}= 
\kappa _{\pm 1}(c_0+c_2)/2$. The analytic solutions (\ref{anali}) of the 
decoupled equations (\ref{gps1}) cannot, however, determine the position 
of individual solitons, which will be fixed in an ad-hoc fashion. 
The energy functional of Eq. (\ref{gps1}) can  now be 
written as 
\begin{align}
 E(\gamma) &=\sum_{j=\pm 1}  \int \kappa_j\Biggr[\frac{1}{2}\left|\frac{d\widehat \phi_{j}}{dx}
 \right|^2
 + \frac{(c_0+c_2)\kappa_j |\widehat \phi_j|^4}{2} 
 \nonumber\\
  &  
   -i\gamma\widehat \phi_{j}^*\frac{d\widehat \phi_{j}}{dx} 
 \Biggr]dx.
  \label{energysep}
\end{align}
For zero magnetization $(m=0)$, using Eq. (\ref{anali}), energy (\ref{energysep})  can be evaluated to yield    
\begin{align} \label{en2si}
E(\gamma) =     -\frac{\gamma^2}{2} -\frac{(c_0+c_2)^2}{96}\approx  -\frac{\gamma^2}{2} - 0.0104167(c_0+c_2)^2.
\end{align}
}

For small values of $\gamma$ ($\gamma <\gamma_c$), the energy of the two-component soliton (\ref{en2si}) is greater than the energy of the three-component soliton (\ref{egysi}), thus making the three-component vector soliton  the ground state. 
The opposite happens for $\gamma > \gamma_c$,  when the phase-separated two-component vector soliton becomes the ground state.  
The crossover takes place when the two minima of energy given by Eqs. (\ref{egysi})  and (\ref{en2si})
are equal, e.g., at  
\begin{equation}\label{crgasi}
 \gamma_c =- {\frac{1}{\sqrt 8}}(c_0+c_2) \approx - 0.353553 (c_0+c_2),
\end{equation}
{For $\gamma <  \gamma_c  $, $E(\gamma)$ of Eq. (\ref{egysi}) is smaller than 
that  of Eq. (\ref{en2si}) making the  overlapping state  the ground state, as will be verified in our numerical calculation. As $\gamma$ increases past $  \gamma_c$, 
for $\gamma>  \gamma_c$ the opposite is true  making the phase-separated state the ground state.  
}

We now show that  the phase separation of the vector soliton demonstrated above is  a consequence of the SO coupling $\gamma p_x \Sigma_z$ used and it is not  a general phenomenon common to other types of the SO coupling. 
This SO coupling is obtained by aligning the electromagnetic fields appropriately \cite{usedspin1/2}.
 In one dimension, for a spin-1 spinor,  there are two other linearly independent SO couplings:  $\gamma p_x \Sigma_x$ and  $\gamma p_x \Sigma_y$, where 
\begin{align}
\Sigma_x= \frac{1}{\sqrt 2} \left( \begin{array}
 {ccccc}
0 & 1 & 0\\
1 & 0 & 1\\
0 & 1 & 0 \end{array} \right), 
\Sigma_y= \frac{i}{\sqrt{2}}\left( \begin{array}
 {ccccc}
0 & -1 & 0\\
1 & 0 & -1\\
0 & 1 & 0 \end{array} \right).
\end{align}
In both cases the SO coupling connects the components $F_z=\pm 1$ with the component $F_z=0$.  Hence when the $F_z=0$ component vanishes, there will be no SO coupling. To illustrate the above claim explicitly 
we  consider the SO coupling $\gamma p_x \Sigma_x$  considered in a previous study \cite{sol1d}. In this case the mean-field GP equation is  \cite{gautam-2}
\begin{align}
 i \frac{\partial \phi_{\pm 1}}{\partial {t}} &=
 \left( -\frac{1}{2}\frac{\partial^2}{\partial  {x}^2}
 +   {c}_0 {\rho}\right)\phi_{\pm 1} {-}  \frac{ {i \gamma}}{\sqrt 2}
  \frac{\partial\phi_{0}}{\partial   x}\nonumber\\ 
 &+c_2( {\rho}_{\pm 1}+ {\rho}_0- {\rho}_{\mp 1})\phi_{\pm 1}+ c_2 \phi_0^2\phi_{\mp 1}^*,\label{gp1}\\
 i\frac{\partial \phi_0}{\partial {t}} &= 
 \left( -\frac{1}{2}\frac{\partial^2}{\partial  {x}^2}
 + {c}_0 {\rho}\right)\phi_0  {\color{red}-}    \frac{ {i \gamma}}{\sqrt 2}\left[  \frac{\partial\phi_{+1}}{\partial   x}
 {+}    \frac{\partial\phi_{-1}}{\partial   x}
 \right]
 \nonumber\\
  &  
  +c_2( {\rho}_{+1}+ {\rho}_{-1})\phi_0+ 2 {c}_2\phi_0^*\phi_{+1}\phi_{-1}.\label{gp3}
\end{align}
With the increase of the SO-coupling strength $\gamma$ it is not possible to have a phase-separated two-component vector soliton of the $F_z=\pm 1$ components only with vanishing $F_z=0$ component, because if we set the $F_z=0$  component   $\phi_0=0$ in  Eqs.  (\ref{gp1}) and (\ref{gp3}),
the SO coupling disappears   and the equations become independent of the SO coupling. The same will be true for the SO coupling $\gamma p_x \Sigma_y$.

\subsection{Moving Soliton}

Although the SO-coupled GP equation is not  Galilean invariant \cite{usedspin1/2},  
 we will show that it is possible to have a moving ferromagnetic soliton of the type considered in this paper, which can propagate maintaining the shape.
Actually the SO coupling terms, and not the nonlinear terms, are responsible for the breakdown and  we consider only the SO coupling terms of Eq. (\ref{gps-1}) as    
\begin{align}\label{sogal}
 i \frac{\partial \phi_{\pm 1}(x,t)}{\partial {t}} &=\left[
  -\frac{1}{2}\frac{\partial^2}{\partial  {x}^2}
 \mp {i {\gamma}}
  \frac{\partial }{\partial   x}\right] \phi_{\pm 1}(x,t) .
\end{align}
Let us consider the Galilean transformation with a velocity $v$ connecting the rest frame to the moving primed frame:
\begin{align}
x'&=x+vt, \quad  t'=t, \\
\frac{\partial }{\partial x}&=\frac {\partial}{\partial x'},\quad  \frac{\partial} {\partial t}=
\frac{\partial}{\partial t'}
+ v \frac{\partial} {\partial x'}. \label{der}
\end{align}
In the absence of SO coupling $(\gamma=0)$, 
 Galilean invariance requires that 
 the form of the Schr\"odinger equation in the primed frame remains unchanged 
provided that the wave functions in the rest and primed frames are related by a phase:
\begin{align}\label{26}
\phi_{\pm 1}(x,t)= e^{i(-vx'+v^2t'/2)}\phi_{\pm 1}'(x',t') 
\end{align}  
which can be proved by a direct substitution of  Eqs. (\ref{der}) and (\ref{26}) into Eq. (\ref{sogal}). 

{
In the presence of SO coupling ($\gamma\ne 0$), 
in the rest frame ($v=0$) 
the solutions of Eq. (\ref{sogal}) are
 \begin{align} \phi_{\pm 1}(x,t)=e^{i(\mp \gamma x +\gamma^2t/2)}. 
\end{align}
In the moving primed frame ($v\ne 0$) the form of Eq. (\ref{sogal}), in the presence of SO coupling $(\gamma \ne 0)$,
 remains unchanged provided that the wave functions in the rest and primed 
frames are related by a phase:
\begin{align}\label{mving}
\phi_{\pm 1}(x,t)&= e^{i(-vx'+v^2t'/2 \pm \gamma v t')}\phi_{\pm 1}'(x',t') ,  
\end{align}
as in Eq. (\ref{26}) for $\gamma =0$, where 
\begin{align}\label{29}
\phi'_{\pm 1}(x',t')&=  e^{i(\mp \gamma x' +\gamma^2t'/2)}. 
\end{align}
A straightforward substitution of Eqs. (\ref{mving}) and (\ref{29}) in Eq. (\ref{sogal})
and  the use of Eq. (\ref{der}) show that the function $\phi_{\pm 1}(x,t) $ of Eq.  (\ref{mving})  is a solution of 
 Eq. (\ref{sogal}) in the rest frame. From Eq. (\ref{mving})  we see that 
apart from the overall phase $(-vx'+v^2t'/2)$, as in Eq. (\ref{26}) for $\gamma =0$, relating the rest and moving frames, there is 
 a shift of time-dependent phase  $\pm \gamma vt'$ for the components $\phi_{\pm 1}$ and a 
shift of phase 0 for the component $\phi_{0}$, which is not affected by the present 
SO coupling.     This phase is not the same for the three components and leads to different energies 
for the three components.  
Hence, although the Galilean invariance is not valid in a strict sense,  the density of the three
components of the  
vector
soliton  
will be conserved 
during motion  for a class of solutions. 
If we include in Eq. (\ref{sogal}), the necessary nonlinear terms to form a localized 
soliton, viz. Eq. (\ref{gps-1}),  this analysis holds provided the added terms do not introduce 
an extra $\gamma$ dependence in the solution, e.g., considering only
solutions of the form $\Phi_{\pm 1}\equiv f(x) \exp(\mp i\gamma x +i\gamma^2t/2) $, where
$f(x)$ is the $\gamma-$independent  spatial profile of the stationary localized wave function. As the function $f(x)$  is independent of $\gamma$, the added nonlinear terms do not interfere in the above 
analysis of Galilean invariabce. 
All ferromagnetic $(c_2<0)$  solitons considered in this paper are of this type, viz. (\ref{Gaussian}) and (\ref{anali}). Hence these solitons are true solitons, which can propagate maintaining shape \cite{sol1d}
or density of individual components. The anti-ferromagnetic $(c_2>0)$ solitons, on the other hand, have a $\gamma-$dependent  spatial profile and cannot propagate maintaining the shape   \cite{sol1d}.  
}

\section{Result and Discussion}
\label{Sec-III}

We numerically solve the coupled partial differential equations  (\ref{gps-1})-(\ref{gps-3}) using the
split-time-step Crank-Nicolson method \cite{Muruganandam} with real- and imaginary-time propagation.
For a numerical simulation there are the FORTRAN \cite{Muruganandam} and C  \cite{cc} programs  and their 
open-multiprocessing \cite{omp} versions  appropriate for using in multi-core processors. 
 The ground 
state is determined by solving  (\ref{gps-1})-(\ref{gps-3}) using 
imaginary time propagation, which neither conserves normalization  nor magnetization. 
Both normalization  and magnetization can be fixed by normalizing  the 
wave-function components appropriately after each time iteration \cite{Bao}.  
The real-time propagation method was used to study the dynamics with the converged solution     
obtained in imaginary-time propagation as the initial state. 
The space and time steps employed in the imaginary-time propagation are $dx =0.05$ and $dt =0.0002$ 
and that in the real-time propagation are  $dx =0.05$ and $dt =0.0001$.

\begin{figure}[!t]
\begin{center}
\includegraphics[trim = 0mm 0mm 0cm 0mm, clip,height=3.5cm,width= 7.5cm,clip]{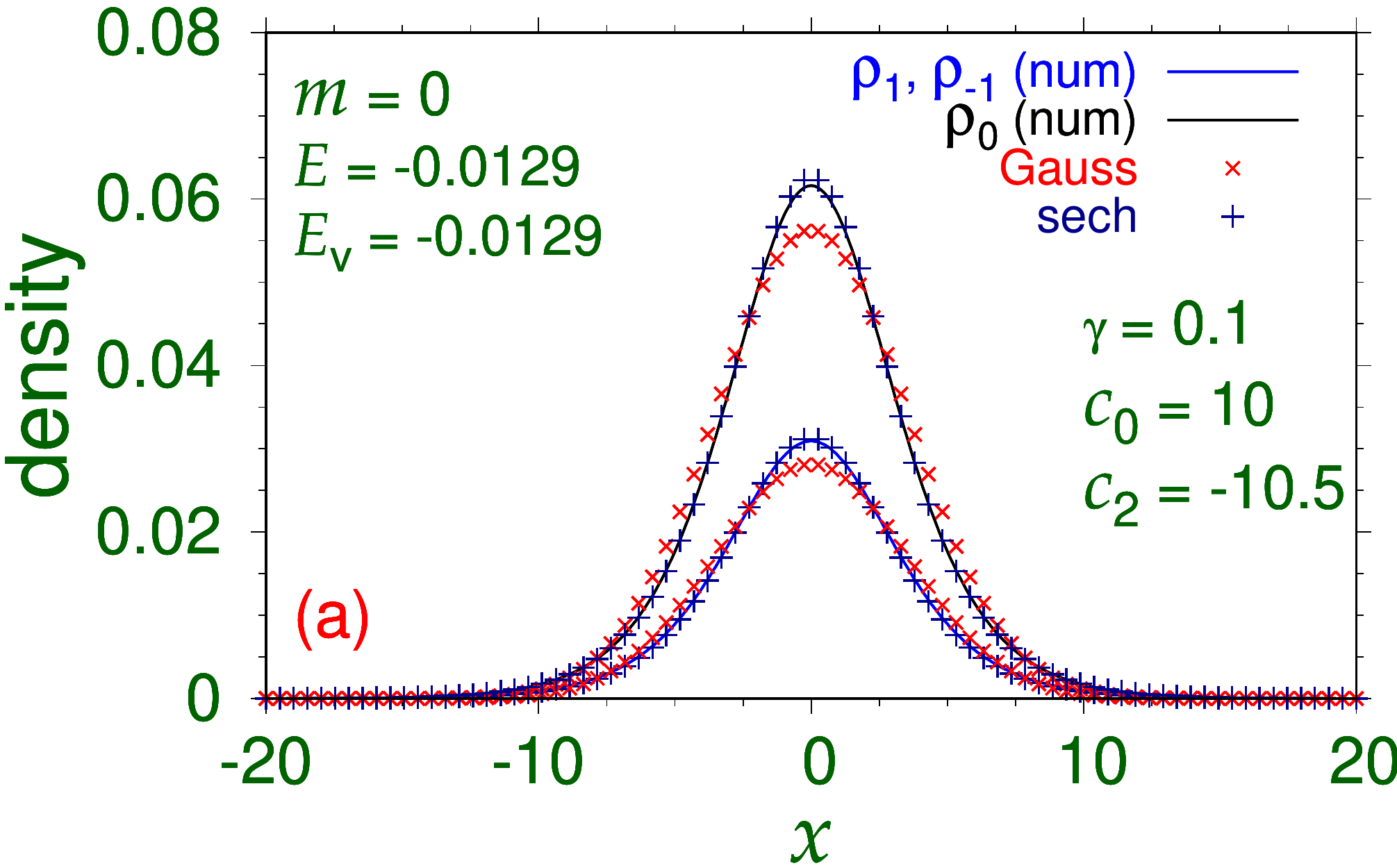}
\includegraphics[trim = 0mm 0mm 0cm 0mm, clip,height=3.5cm,width= 7.5cm,clip]{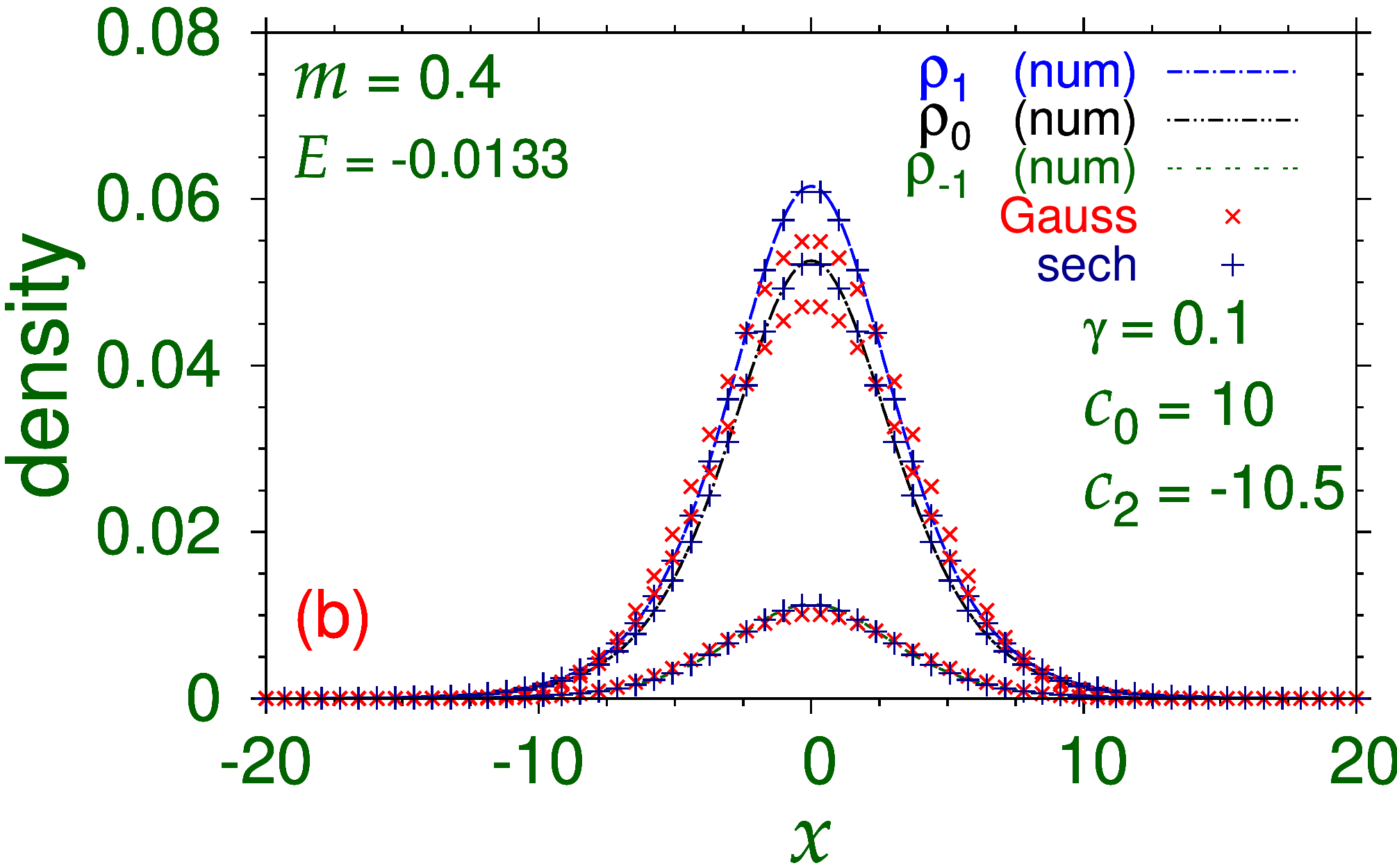}
\includegraphics[trim = 0mm 0mm 0cm 0mm, clip,height=3.5cm,width= 7.5cm,clip]{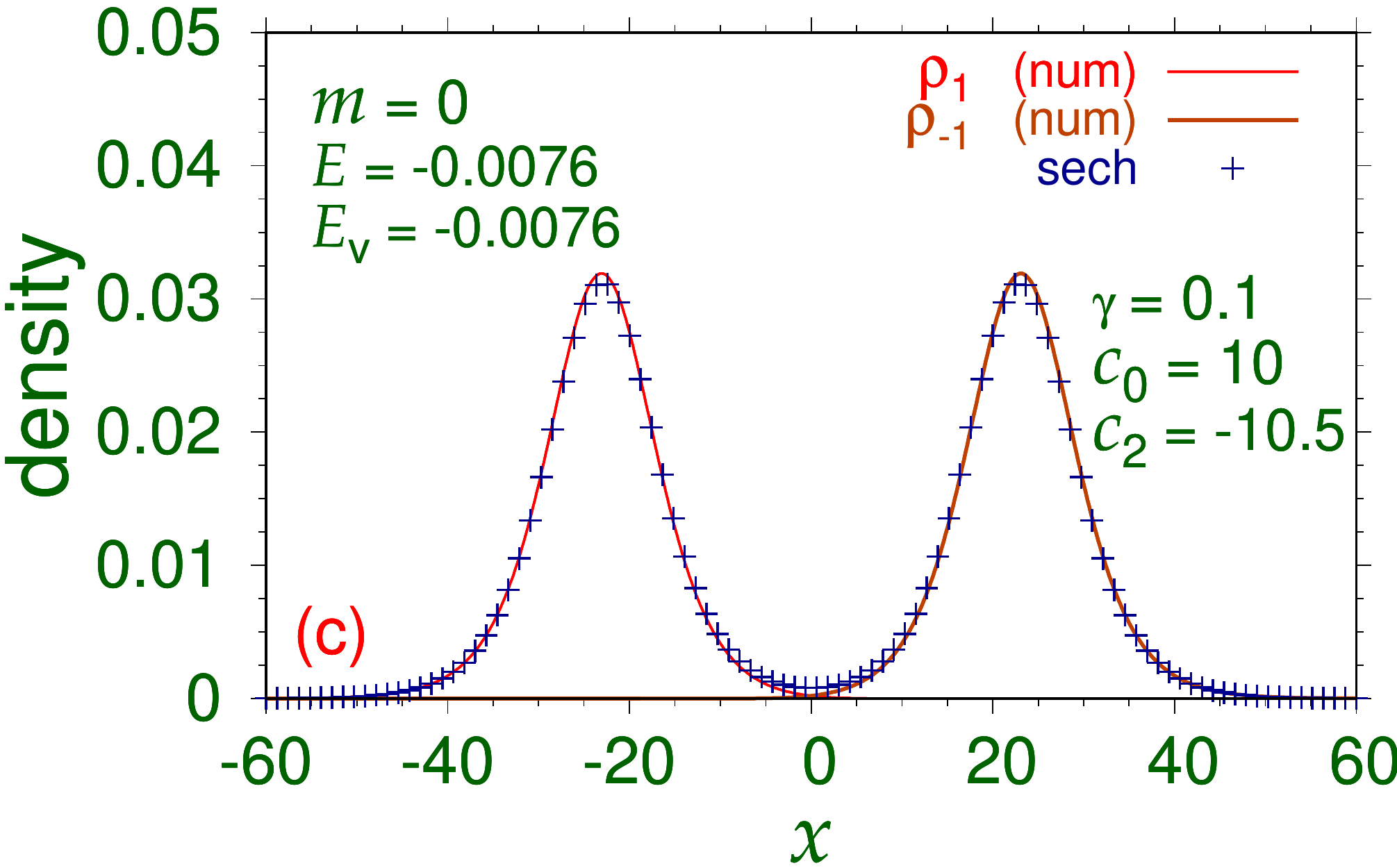}
\includegraphics[trim = 0mm 0mm 0cm 0mm, clip,height=3.5cm,width= 7.5cm,clip]{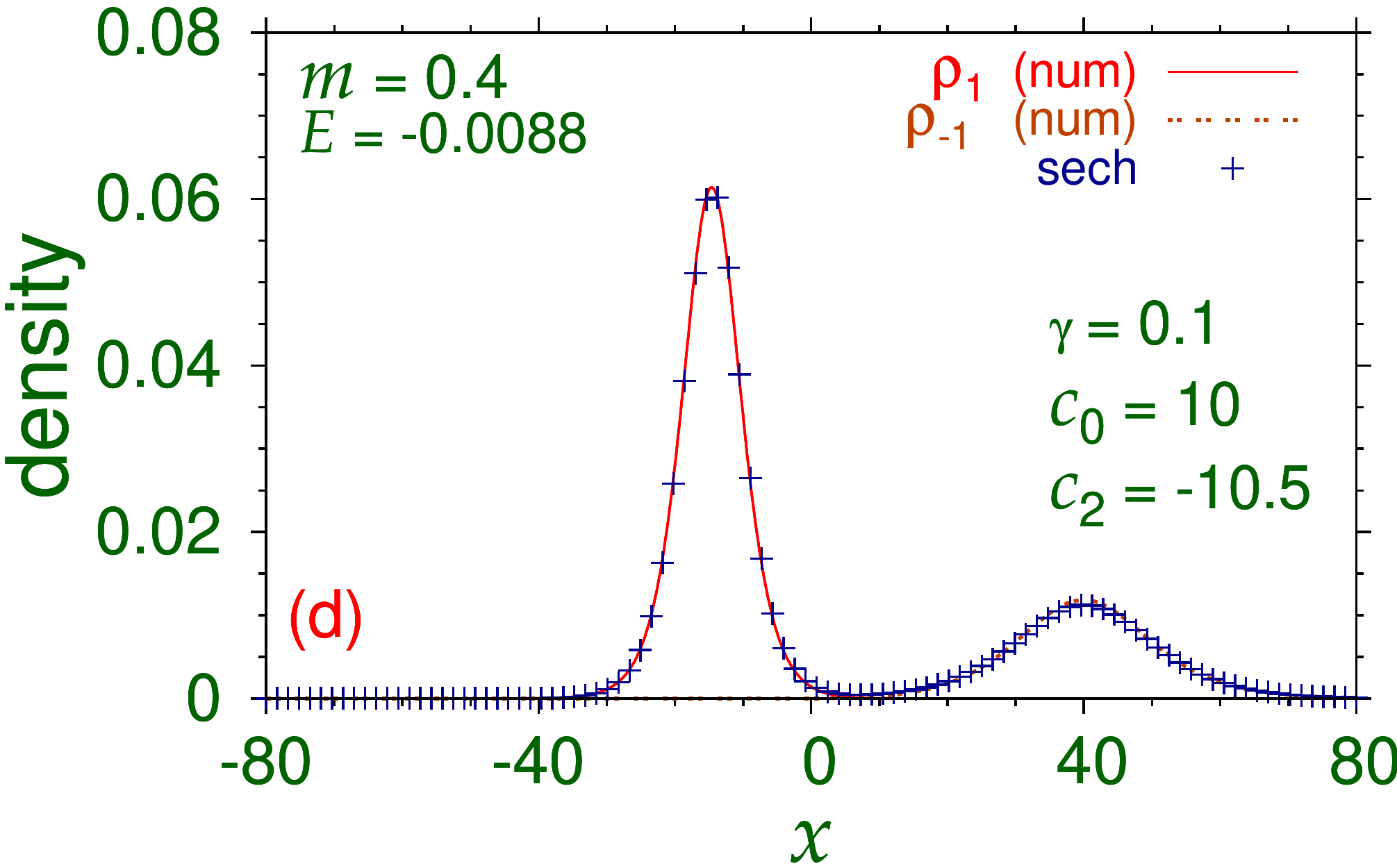}

\caption{(Color online)   Numerical (num) and analytic   densities obtained with 
Gaussian (Gauss) and hyperbolic secant  (sech) functions 
$\rho_j, j=0, \pm 1$ 
of the three components for the lowest-energy overlapping ground-state  vector  soliton 
 with $c_0 = 10$, $c_2 = -10.5$, and  $\gamma=0.1$  ($\gamma  < \gamma_c = 0.1768)$  for (a) $m=0,$ and  (b)    $m=0.4$. 
 The same for the phase-separated  two-component vector soliton in an excited state are 
shown in (c)  and (d),  respectively. { The soliton profiles remain unchanged with a variation of $\gamma$ in the domain $0.1  > \gamma >0$}. Numerical ($E$) energy and variational ($E_v$) energy from $\text{sech}$ ansatz are shown.
 The numerical energies  displayed in  plots   (c) and (d) are larger than those in (a) and (b), respectively. 
All quantities in this and following figures  are dimensionless.}
\label{fig1} \end{center}
\end{figure}

\subsection{Stationary solitons and their stability}
 
The initial wave function in imaginary-time propagation is taken as the variational function (\ref{Gaussian})
with the Gaussian form for the function $\phi(x)$ and with $\gamma =0$.  To study the phase separation of the components of the vector 
soliton efficiently, it is appropriate to give a small separation between the positions of the $F_z=\pm 1$
initial state functions while  maintaining $F_z=0$ component at the origin as 
\begin{equation}
\Phi\equiv  \left( \begin{array}{c} \phi_{+1}\\
\phi_0\\
\phi_{-1}
 \end{array} \right)
= \frac{1}{2}\left( \begin{array}{c}
(1+m)\phi(x+a) \\
\sqrt{2(1-m^2)} \phi(x)\\
(1-m)\phi(x-a) \end{array} \right),  
\label{spnrs}
\end{equation}
where $\phi$ is the normalized Gaussian function (\ref{gauss}) and 
$a$ is a small number. If $\gamma < \gamma_c $,  the critical value for phase separation, 
the $F_z=\pm 1$ components move to the center to form a fully overlapping three-component vector soliton 
in the final converged configuration. However, if $\gamma > \gamma_c$, the  $F_z=\pm 1$ components move outwards 
to form a fully separated two-component vector soliton with the vanishing of the $F_z=0$ component.  The imaginary-time propagation method prefers  to maintain the symmetry of the initial state: overlapping or separated. By taking $a=0$ in Eq. (\ref{spnrs}) it is possible to find the overlapping excited state for $\gamma > \gamma_c$, where the ground state is phase separated; also by taking a large value of $a$    it is possible to find the  phase-separated state for 
 $\gamma < \gamma_c$,  where the ground state is overlapping.

\begin{figure}[!t]
\begin{center}
 
\includegraphics[trim = 0mm 0mm 0cm 0mm, clip,height=3.5cm,width= 7.5cm,clip]{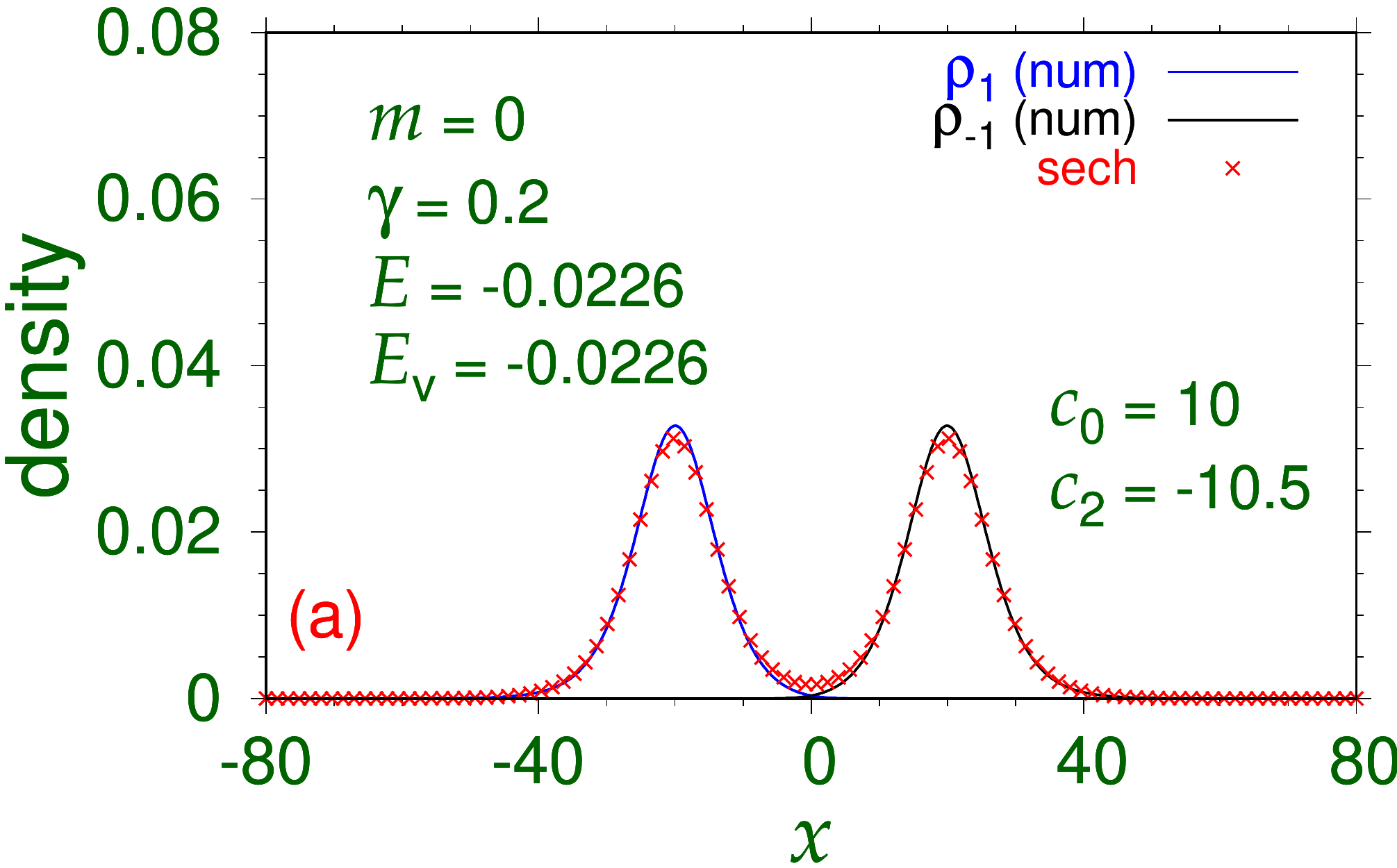}
\includegraphics[trim = 0mm 0mm 0cm 0mm, clip,height=3.5cm,width= 7.5cm,clip]{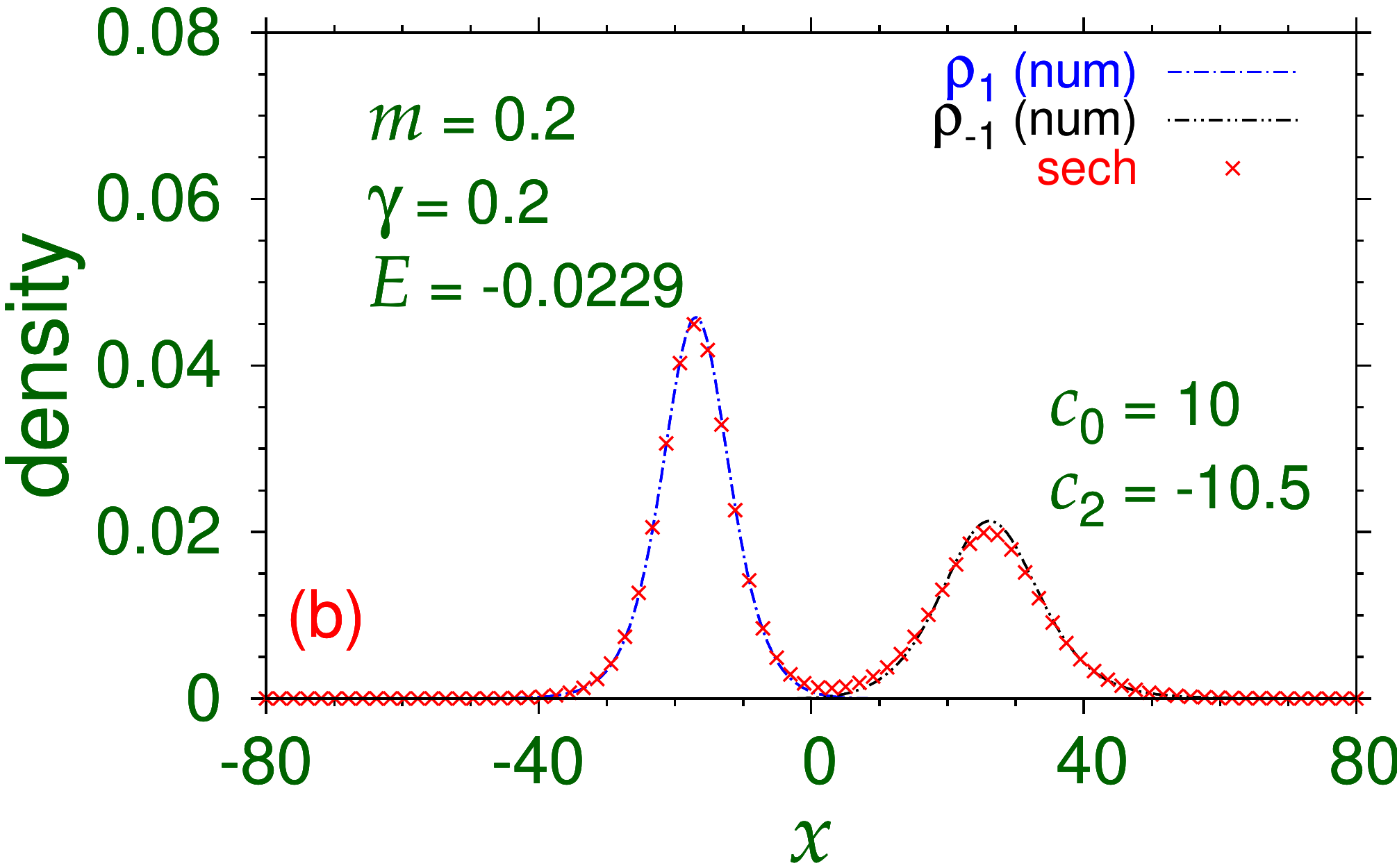}
\includegraphics[trim = 0mm 0mm 0cm 0mm, clip,height=3.5cm,width= 7.5cm,clip]{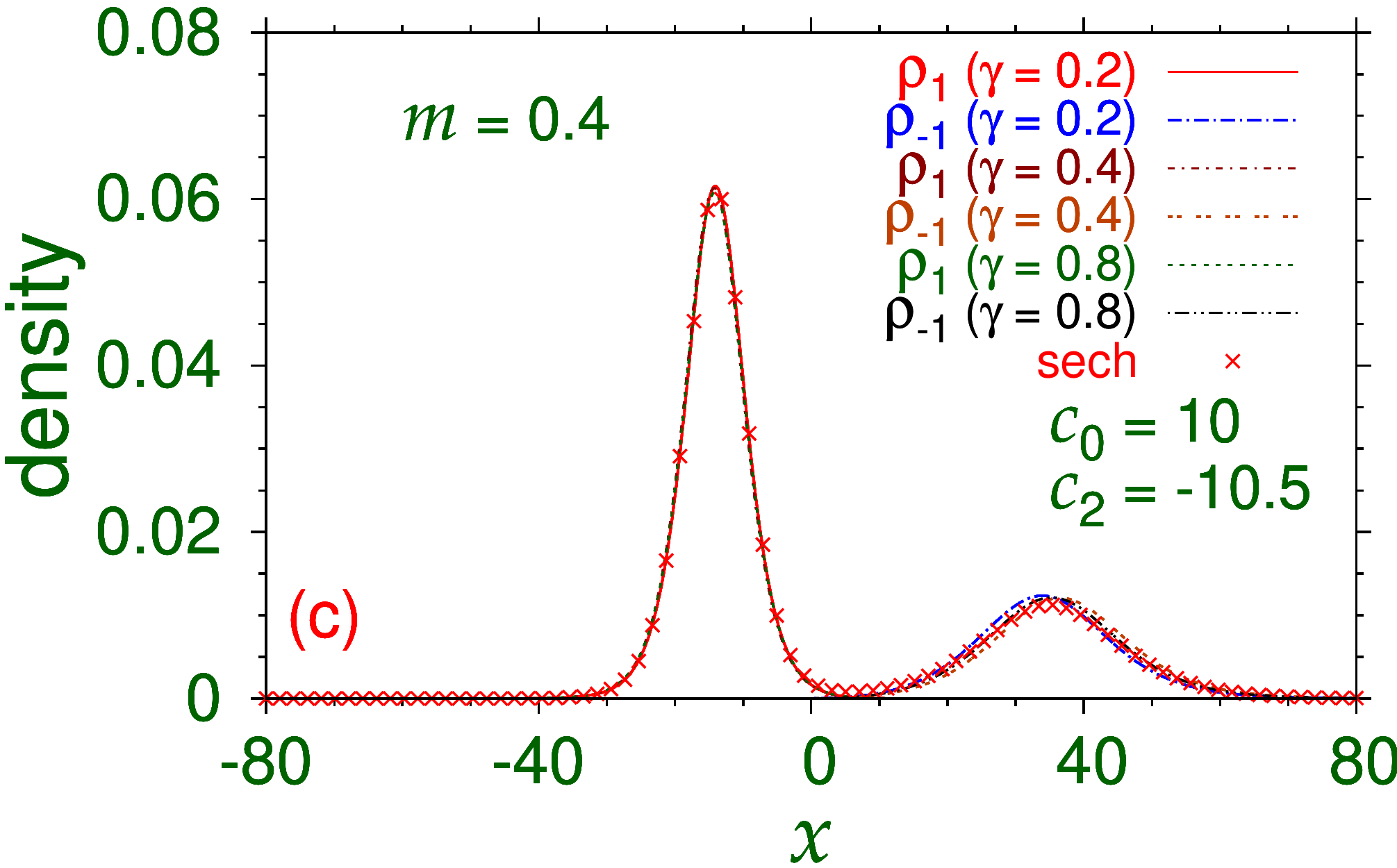}

\caption{(Color online)    Numerical (num) and analytic   densities 
with the hyperbolic secant  function (sech) $\rho_j, j= \pm 1$ 
of the two components for the lowest-energy phase-separated   vector  soliton 
 with $c_0 = 10$, $c_2 = -10.5$, $\gamma =0.2 > \gamma_c = 0.1768$  for (a) $m=0,$ and  (b)    $m=0.2$, and  (c) $m=0.4$.  Numerical ($E$) energy and variational ($E_v$) energy from $\text{sech}$ ansatz are shown.
   }

\label{fig2} \end{center}

\end{figure}

We perform our calculation with the parameters $c_0=10$   and $c_2= -10.5$ (ferromagnetic), so that $c_0+c_2=-0.5< 0$, 
to make the system attractive to have a vector soliton. { The variational approximation  in  Sec. \ref{II} demonstrates  that the present soliton profiles are determined entirely by the combination $(c_0+c_2)$ of the interaction strengths $c_0$ and $c_2$, viz. Eqs. (\ref{minal}) and (\ref{minalsi}), which is also confirmed by our numerical calculation.  Hence without losing generality, we consider in this section only positive values of $c_0$ in such a way that 
$c_0+c_2< 0$.}
 In this case the critical SO-coupling strength  for phase separation (\ref{crgasi})  is  $\gamma_c =  0.1768 $. In Figs. \ref{fig1}(a)-(b) we display the density of the components of the minimum-energy overlapping ground-state vector soliton for $\gamma =0.1$ $(\gamma< \gamma_c)$ and for magnetization $m=0$ and  0.6, respectively. The result  of the analytic variational approximation with the Gaussian and
hyperbolic secant    functions is also displayed in these plots.  The variational results for the width $\alpha$ and $\sigma$ given by (\ref{minal}) 
and (\ref{minalsi}) are independent 
of magnetization $m$ and SO-coupling $\gamma$ and are the same for all components. The same is found to be true in the numerical calculation, in good agreement with the analytic approximation.  In Figs. \ref{fig1}(c)-(d) we plot the densities of the phase-separated two-component vector soliton for the same parameters as in (a)-(b), respectively. In the case of the phase-separated two-component vector solitons,  the analytic  result cannot determine the positions of the  
component solitons,  which have been introduced arbitrarily to fit the position of the components. In plots Fig. \ref{fig1}(a)-(d), the soliton profiles are practically unchanged for $0<\gamma <0.1$, although the energy is changing, viz. Eqs. (\ref{egy}) and (\ref{egysi}). In Figs. \ref{fig1}(a)-(b)  we find that the analytic results obtained
with the hyperbolic secant  function are superior to those obtained with the Gaussian function. Hence in the following we will only show the analytic results obtained with the hyperbolic secant   function.

\begin{figure}[!t]
\begin{center}
\includegraphics[trim = 0mm 0mm 0cm 0mm, clip,height=3.5cm,width= 7.5cm,clip]{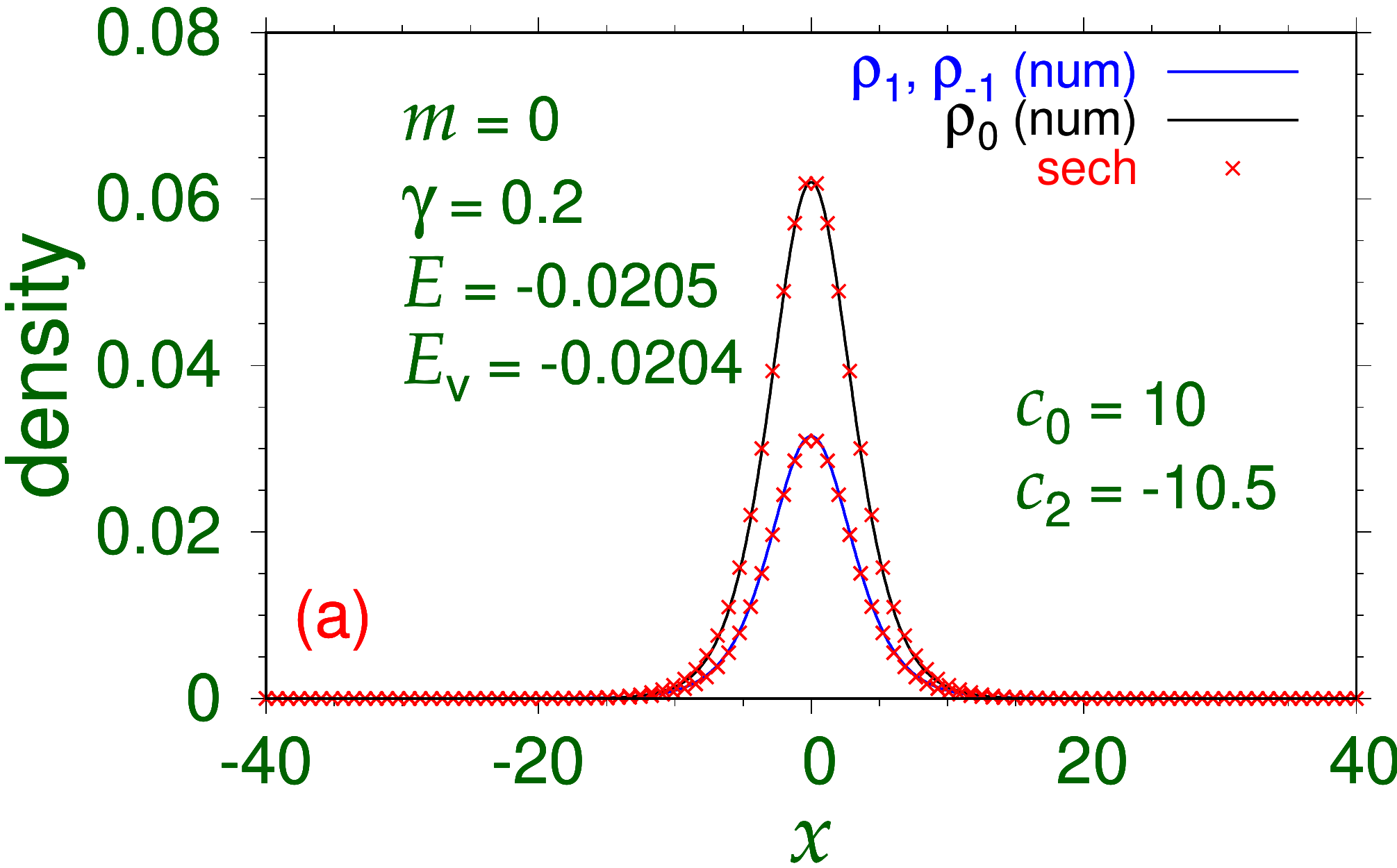}
\includegraphics[trim = 0mm 0mm 0cm 0mm, clip,height=3.5cm,width= 7.5cm,clip]{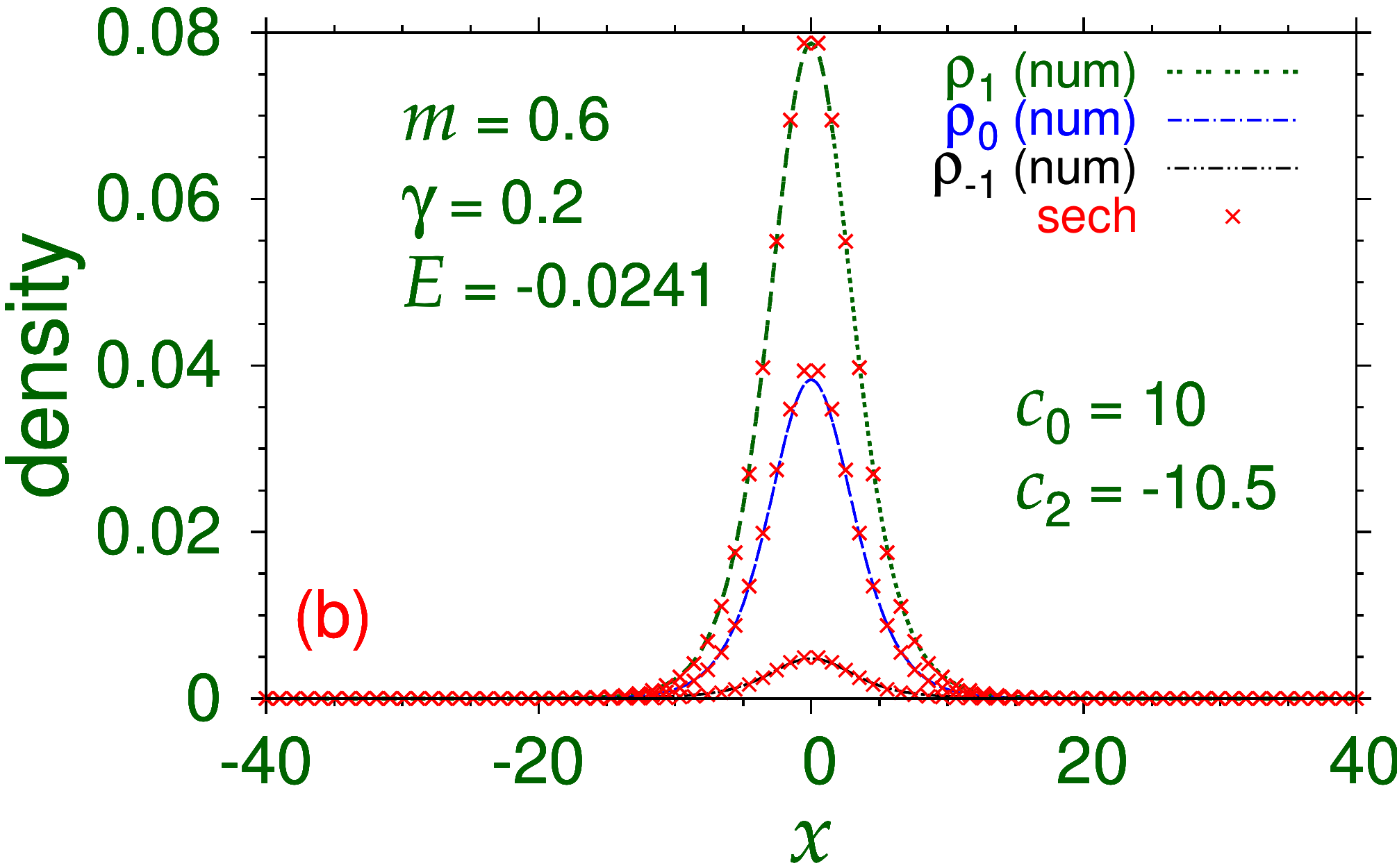}

\caption{(Color online)   Numerical (num)  and analytic   densities with the hyperbolic secant 
  function (sech)
$\rho_j, j=0,  \pm 1$ 
of the three components for the overlapping    vector  soliton in the excited state  
 with $c_0 = 10$, $c_2 = -10.5$, $\gamma =0.2 > \gamma_c = 0.1768$  for (a) $m=0,$ and  (b)    $m=0.6$.      Numerical ($E$) energy and variational ($E_v$) energy from $\text{sech}$ ansatz are shown. }
\label{fig3} \end{center}
\end{figure}

\begin{figure}[!t]
\begin{center}
\includegraphics[trim = 0mm 0mm 0cm 0mm, clip,height=3.6cm,width= 7.5cm,clip]{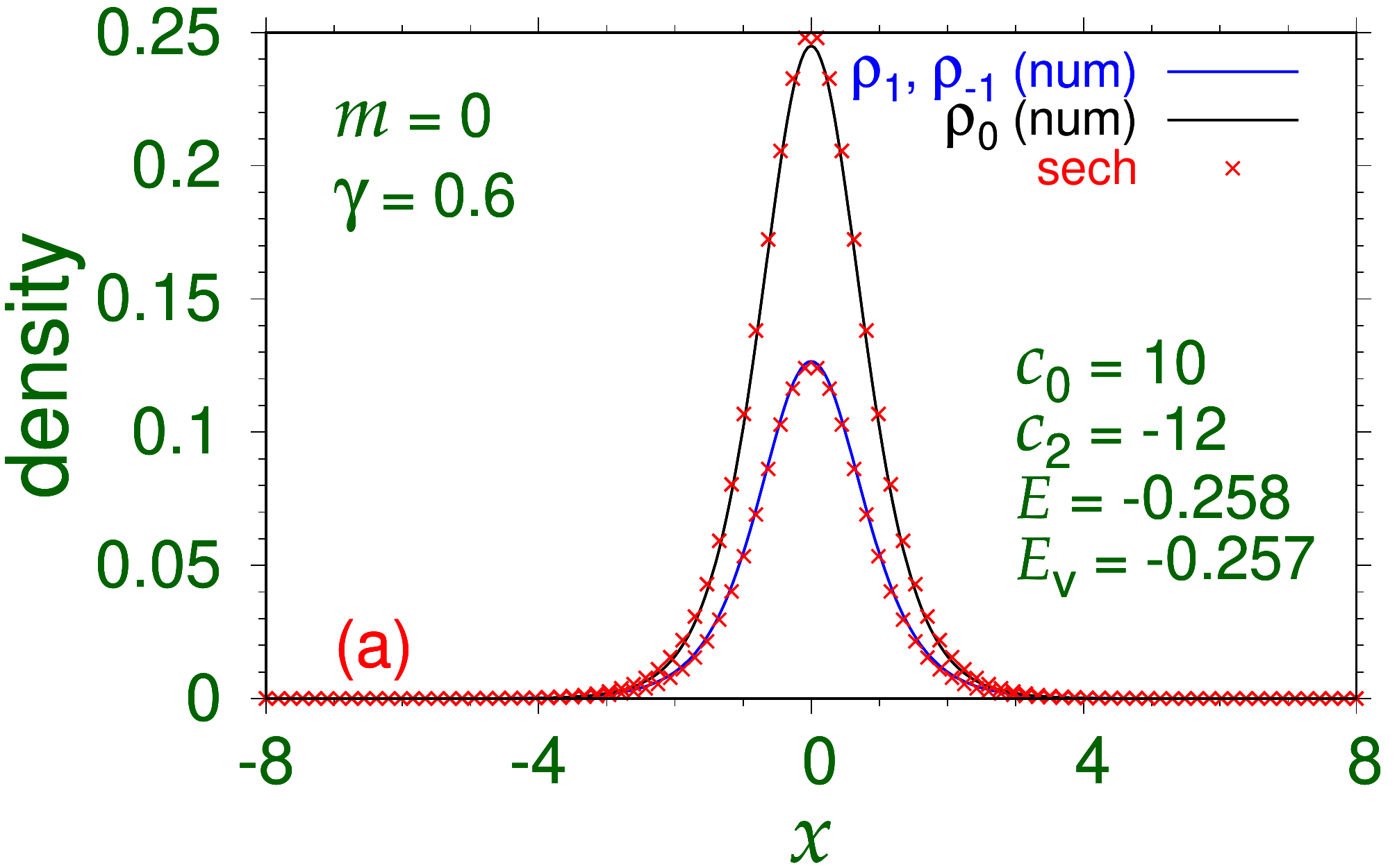}
\includegraphics[trim = 0mm 0mm 0cm 0mm, clip,height=3.6cm,width= 7.5cm,clip]{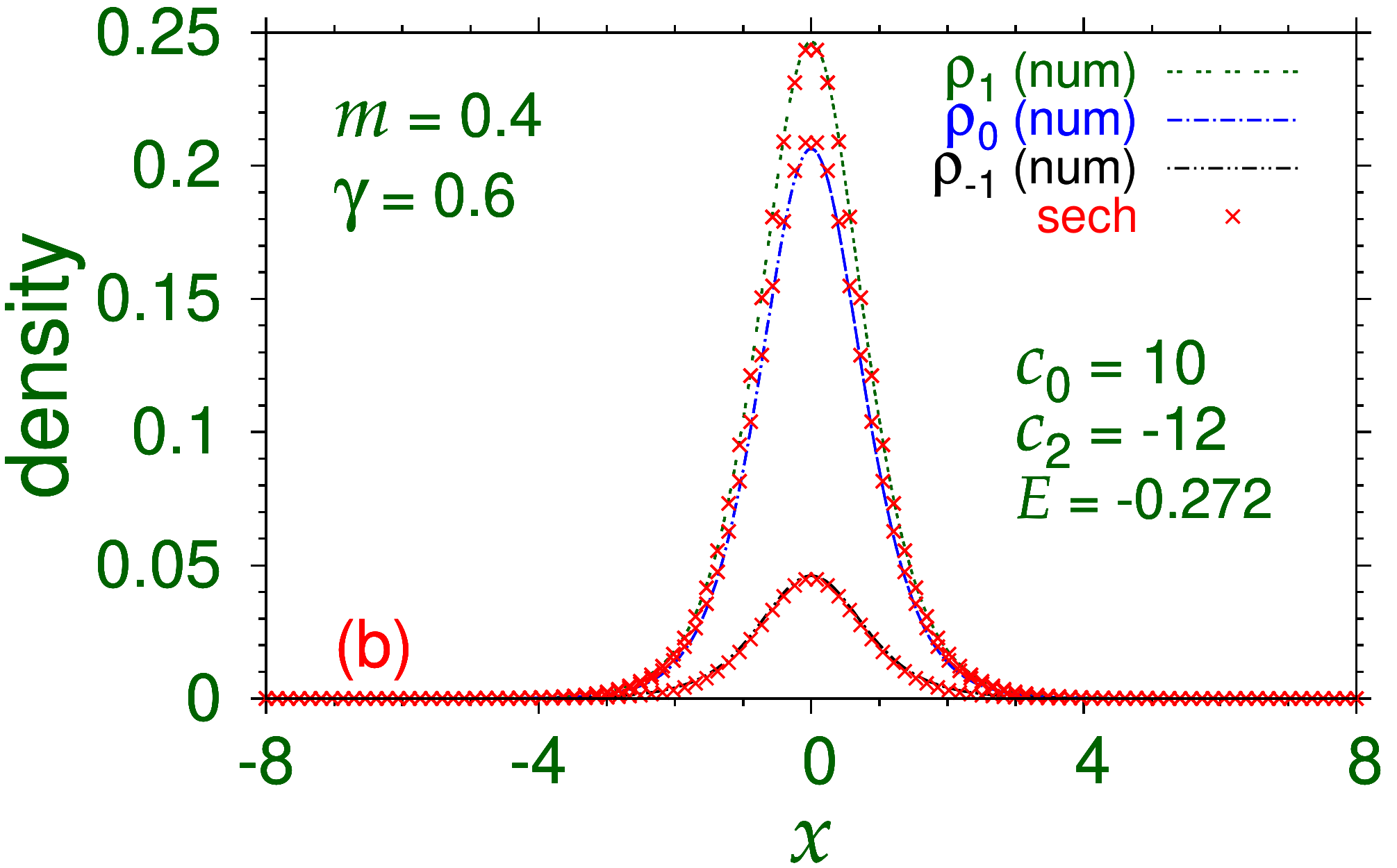}
\includegraphics[trim = 0mm 0mm 0cm 0mm, clip,height=3.6cm,width= 7.5cm,clip]{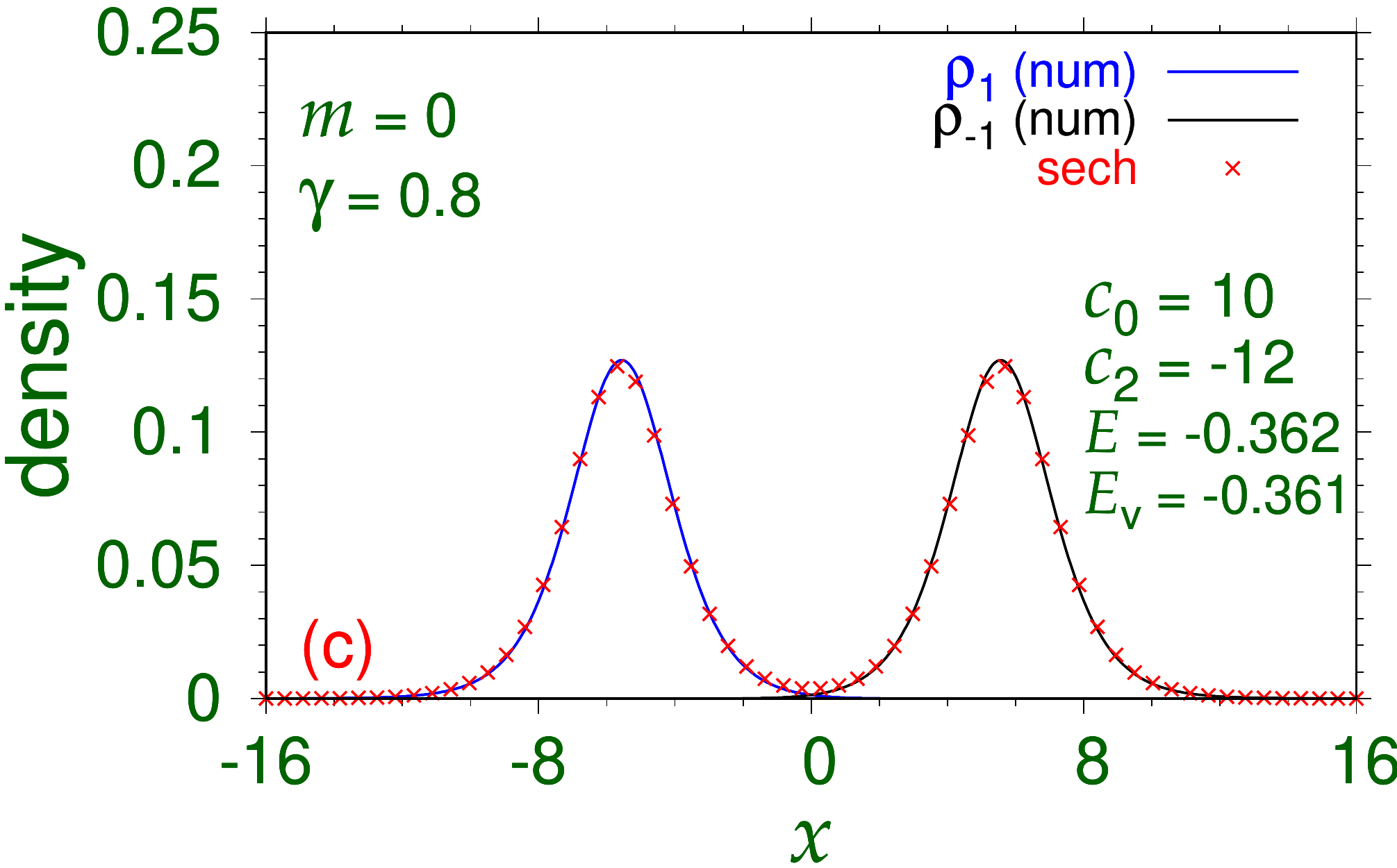}
\includegraphics[trim = 0mm 0mm 0cm 0mm, clip,height=3.6cm,width= 7.5cm,clip]{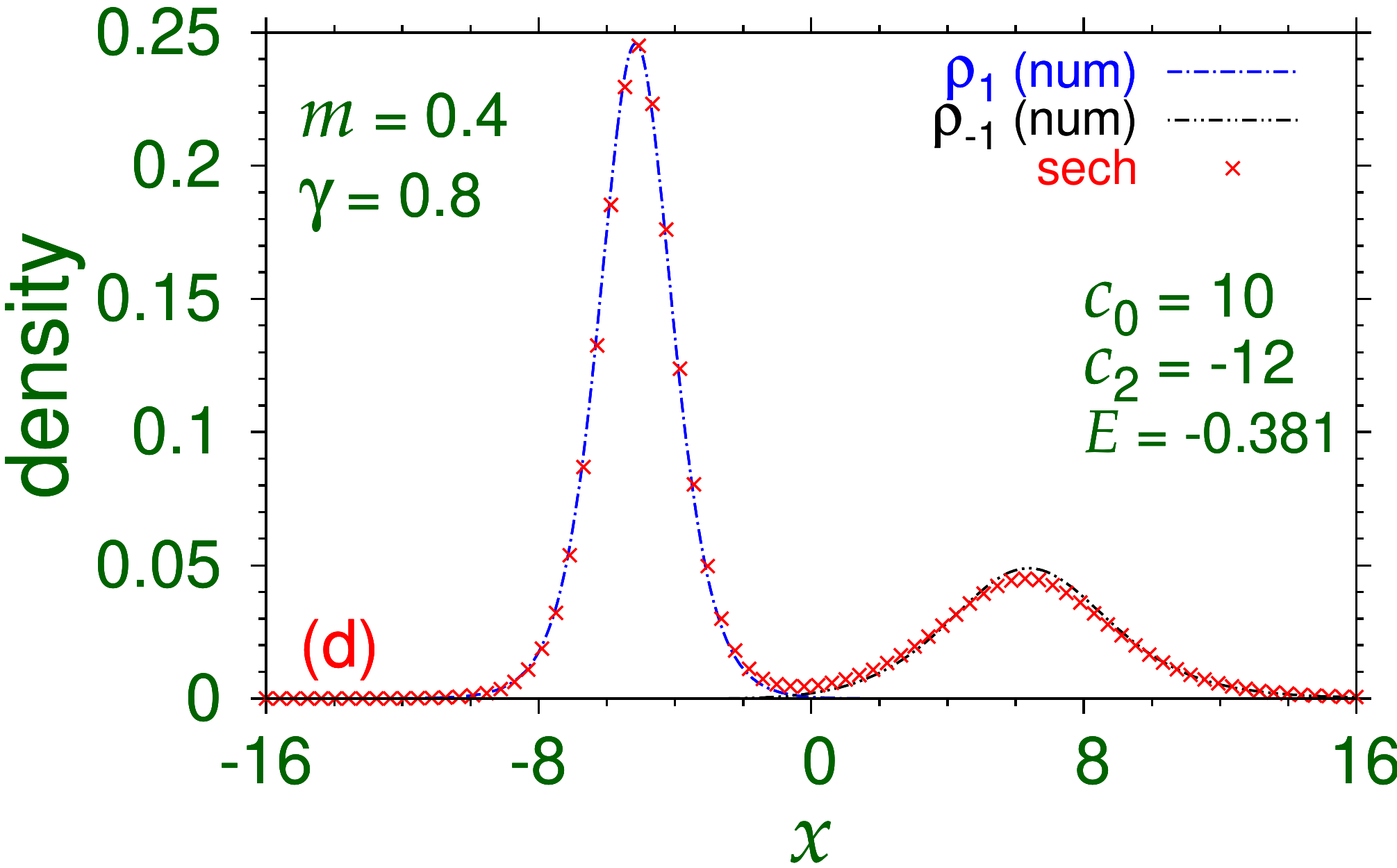}

\caption{(Color online)   Numerical (num) and analytic   densities 
with the hyperbolic secant   function (sech) $\rho_j, j=0,  \pm 1$ 
of the three components of  the overlapping    vector  soliton in the ground  state  
 with $c_0 = 10$, $c_2 = -12$, $\gamma =0.6 < \gamma_c = 0.7071$  for (a) $m=0,$ and  (b)    $m=0.4$. The same 
for the two components  $\rho_j, j=  \pm 1$    of  the phase-separated   vector  soliton in the ground  state 
for  $\gamma = 0.8 >\gamma_c$  and  (c)   $m=0,$ and  (d)    $m=0.4$. 
 Numerical ($E$) energy and variational ($E_v$) energy from $\text{sech}$ ansatz are shown.}
\label{fig4} \end{center}
\end{figure}

In Figs. \ref{fig2} we plot the densities of the phase-separated two-component  minimum-energy ground-state vector soliton for  $c_0=10,$ $c_2= -10.5$, and $\gamma =0.2 >\gamma_c=0.1768$  for (a) $m=0,$ (b)   $m=0.2,$ and  (c)   $m=0.4.$ 
and compare these with the analytic   result from the hyperbolic secant   function.  To show that the density profiles for the phase-separated solitons are
practically  independent of the SO-coupling strength $\gamma $ ($\gamma>\gamma_c$), as predicted by the analytic relation (\ref{anali}), we exhibit in Fig. \ref{fig2}(c) the results for   $\gamma = 0.2,0.4$ and 0.8 in good agreement with each other.

\begin{figure}[!t]
\begin{center}
\includegraphics[trim = 0mm 0mm 0cm 0mm, clip,width=.88\linewidth,clip]{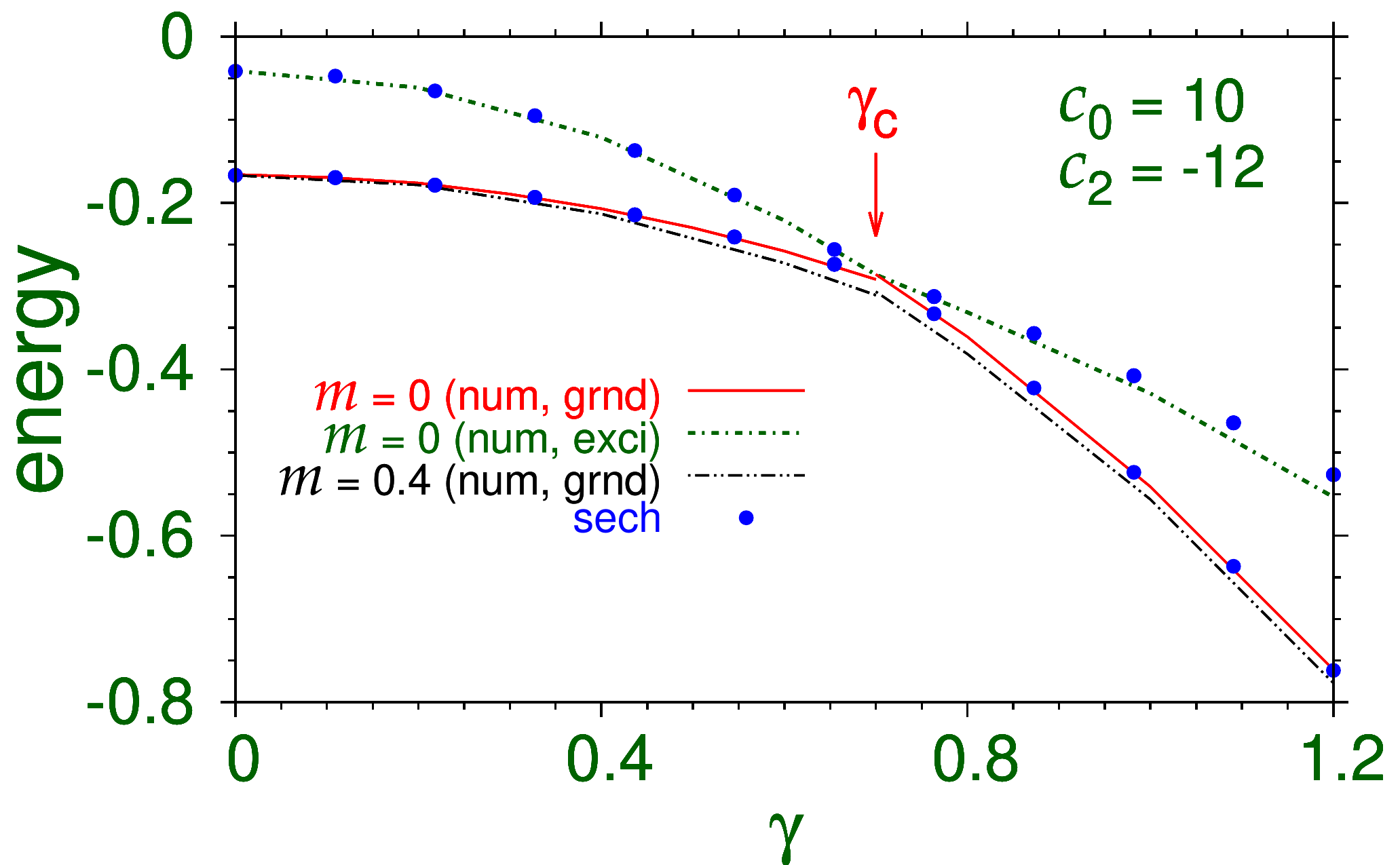}

\caption{(Color online) Numerical (num)    results of energy of $m=0$ ground- (grnd) and excited- (exci) state vector solitons   versus
 SO-coupling 
strength $\gamma$, for interaction strengths $c_0=10$ and $c_2 =-12$,
  displayed by  full and dashed lines, respectively.  The numerical results for for the $m=0.4$ ground-state vector solitons are shown by dashed-dotted line.  The analytic     results (sech) obtained with the hyperbolic secant   function  for the ground and excited states are displayed as solid circles.  The analytic result is independent of $m$ and hence is the same for $m=0$ and $m=0.4$  vector solitons.  
}
\label{fig5} \end{center}
\end{figure}

For $\gamma>\gamma_c$, the phase-separated two-component vector solitons are  the ground states. while  the overlapping three-component solitons become excited states.  In Figs. \ref{fig3} we exhibit the density profiles of these excited states  
for (a)  $m=0$ and (b) $m=0.6$, respectively, and for $\gamma=0.2, c_0=10, c_2=10.5$.
  Although, these are  excited states, the analytic results for  energies and widths  are in good agreement with the numerical energies.

To  show the nature of the solitons with increased attraction we next consider $c_0=10, c_2 =-12$ corresponding to a { net attraction}  
$c_0+c_2 =-2$.     As this net attraction increases, the width of the soliton reduces rapidly, viz. Eqs. (\ref{minal}) and (\ref{minalsi}), whereas the critical $\gamma_c$,
viz. Eq. (\ref{crgasi}), increases.  This is why we did not consider a larger value of $|c_0+c_2|$. In this case the critical  SO coupling  $\gamma_c=0.7071$. We display    the  densities of the 
overlapping ground states 
in Figs. \ref{fig4} for (a) $m=0, \gamma =0.6$  and   (b)   $m=0.4, \gamma =0.6$  and compare these with the analytic  counterparts. These densities remain practically unchanged for all $\gamma <\gamma_c$. For $\gamma >\gamma_c=0.7071$, the overlapping states become excited states and the densities of the phase-separated two-component ground states for $\gamma =0.8$  are shown in Fig. \ref{fig4} for (c) $m=0$  and (d)  $m=0.4$. The analytic results are found to be in good agreement with the numerical calculation.

We now study the evolution of the energy of ground- and excited-state vector solitons as a function of the  SO-coupling strength $\gamma$,
for interaction strengths $c_0 =10, c_2 =-12$.
  In Fig. \ref{fig5},  we display  the numerical results for energy of the $m=0$  ground-state  (full line)
and excited-state  (dashed line) vector solitons. 
 For $\gamma < \gamma_c$, the critical SO-coupling strength for the formation of 
phase-separated two-component ground-state vector solitons, the ground-state solitons are the overlapping three-component solitons. 
For $\gamma > \gamma_c$,  the ground state solitons are the phase-separated  two-component solitons.
 The numerical results for the $m=0.4$ vector solitons are also displayed in this figure as dashed-dotted line.
The analytic   results for energy are independent of $m$ and are displayed by solid circles.

\begin{figure}[!t]
\begin{center}
\includegraphics[trim = 0mm 0mm 0cm 0mm, clip,width=.8\linewidth,clip]{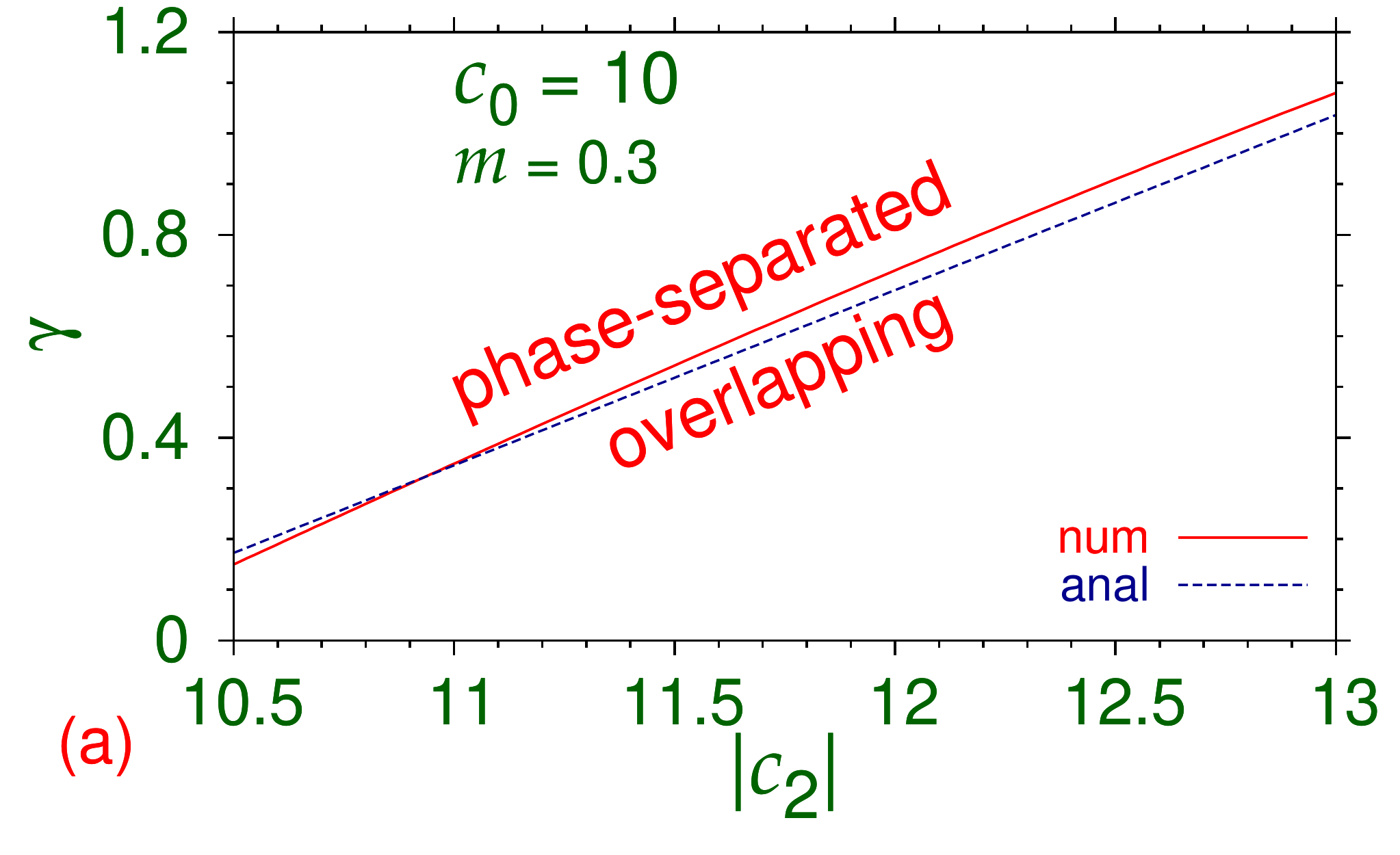}
\includegraphics[trim = 0mm 0mm 0cm 0mm, clip,width=.8\linewidth,clip]{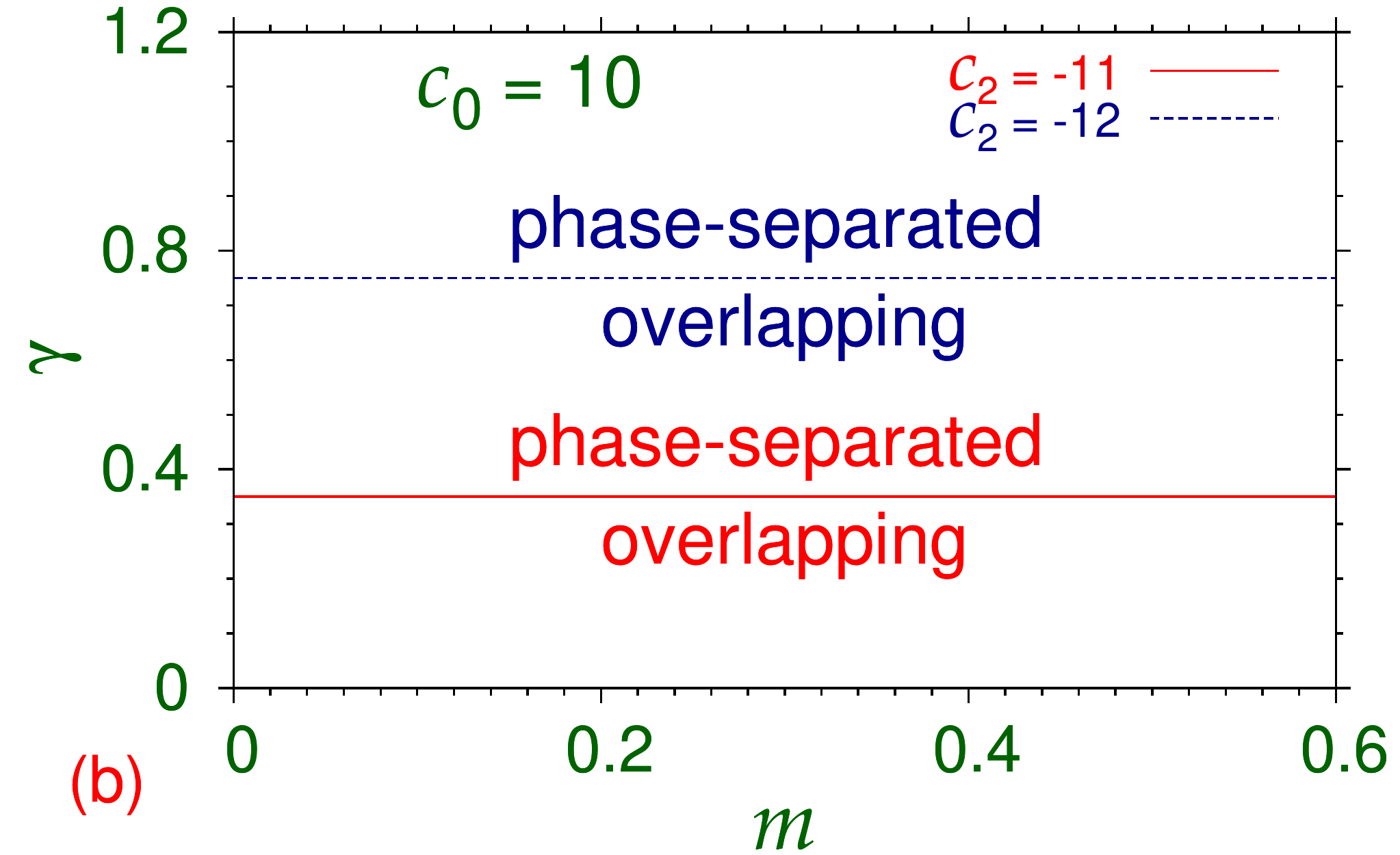}
 
\caption{(Color online)  (a) The numerical and  analytic  phase plot of  phase separation in the
$\gamma-|c_2|$  parameter space for $c_0=10$ and $m=0.3$. (b) The same in the $\gamma-m$ parameter space
for $c_0=10$ and $c_2=-11, -12$. In this case the numerical and  analytic results are very close to each other and only the former is shown.}
\label{fig6} \end{center}
\end{figure}

The phase separation of the three-component vector soliton  in parameter space is illustrated next 
for interaction strength $c_2$ and magnetization $m$ with the variation of  SO-coupling strength $\gamma$.
The analytic  result depends on the interaction-strength combination $c_0+c_2$, whereas the 
numerical result should  depend on both $c_0$ and $c_2$.  In Fig. \ref{fig6}(a) we show the phase separation 
in the $\gamma-|c_2|$ parameter space for interaction strength $c_0=10$ and magnetization $m=0.3$. Both numerical and 
analytic  results, in close agreement with each other, are shown. In Fig. \ref{fig6}(b) the 
numerical results of phase separation are illustrated in the  $\gamma-m$  parameter space for $c_0=10$
and for $c_2=-11$ and -12.

\begin{figure}[!t]
\begin{center}
\includegraphics[trim = 0mm 0mm 0cm 0mm, clip,width=\linewidth,clip]{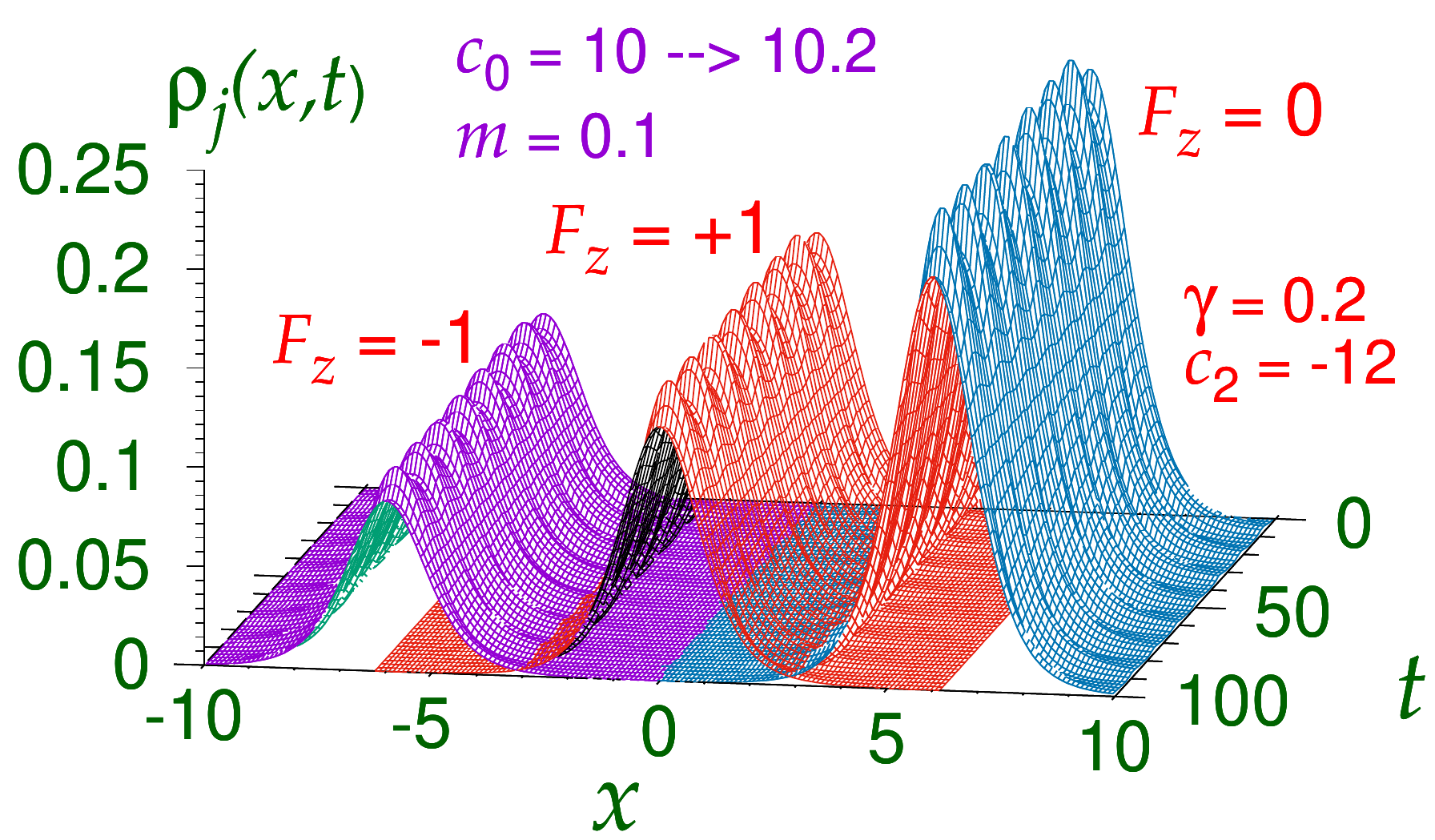}

\caption{(Color online) Density profile of a vector soliton during real-time propagation demonstrating its dynamical stability.
For a better view of the individual components, the component densities are plotted after a spatial displacement among these. 
The initial wave function was obtained by imaginary-time simulation with parameters 
$c_0=10, c_2=-12, m=0.1, \gamma=0.2$ and the real-time propagation was executed after 
changing $c_0$ to 10.2.
 }

\label{fig7} \end{center}
\end{figure}

\subsection{Dynamical stability and phase separation}

To demonstrate that the vector soliton is dynamically stable we subject the ground-state vector soliton profile, 
obtained by imaginary-time simulation, to real-time propagation for a long time after giving a perturbation by 
changing the { interaction strength $c_0$ slightly at time $t=0$. The profile of the vector soliton is very sensitive to $c_0$, viz. (\ref{minal}) and (\ref{minalsi}). 
 }  For this purpose, we consider the overlapping 
 vector soliton 
obtained with parameters $c_0=10, c_2=-12, m=0.1, \gamma =0.2$. The real-time propagation during 100 time units 
for this soliton was 
executed upon  {changing the interaction strength $c_0 $ from 10 to 10.2 at time $t=0$.}     In Fig. \ref{fig7}  we exhibit the density profile of the three 
components of the vector soliton during real-time propagation. For a better view, we have displaced the density profile 
of components $F_z=-1$ and $F_z=0$ to $x=-5$ and $x=+5$, respectively, leaving the  $F_z=+1$ component at $x=0$. The 
long-time stable propagation of the components of the vector soliton establishes its dynamical stability.

\begin{figure}[!t]
\begin{center}
\includegraphics[trim = 0mm 0mm 0cm 0mm, clip,width=.9\linewidth,clip]{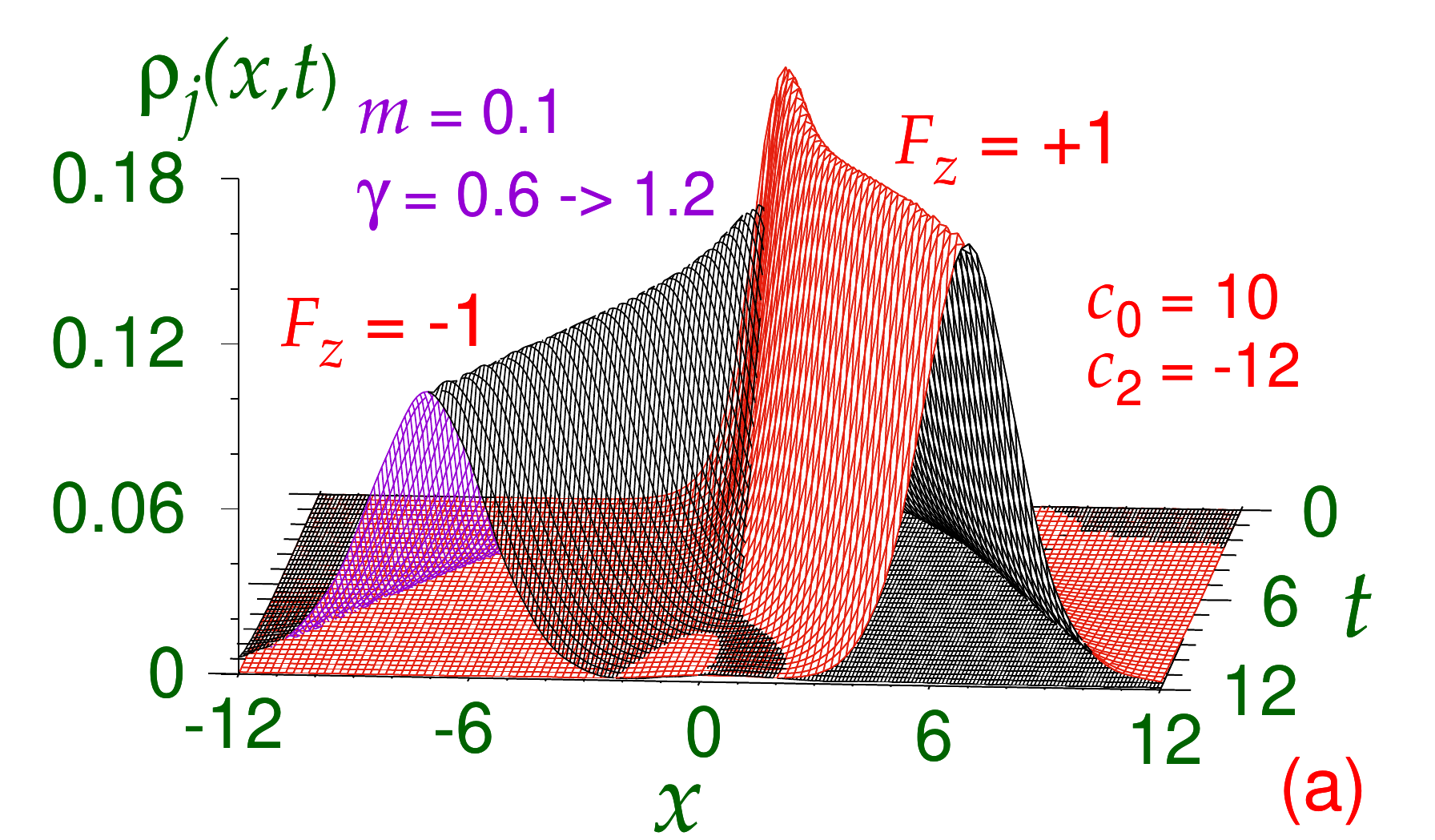}
\includegraphics[trim = 0mm 0mm 0cm 0mm, clip,width=.9\linewidth,clip]{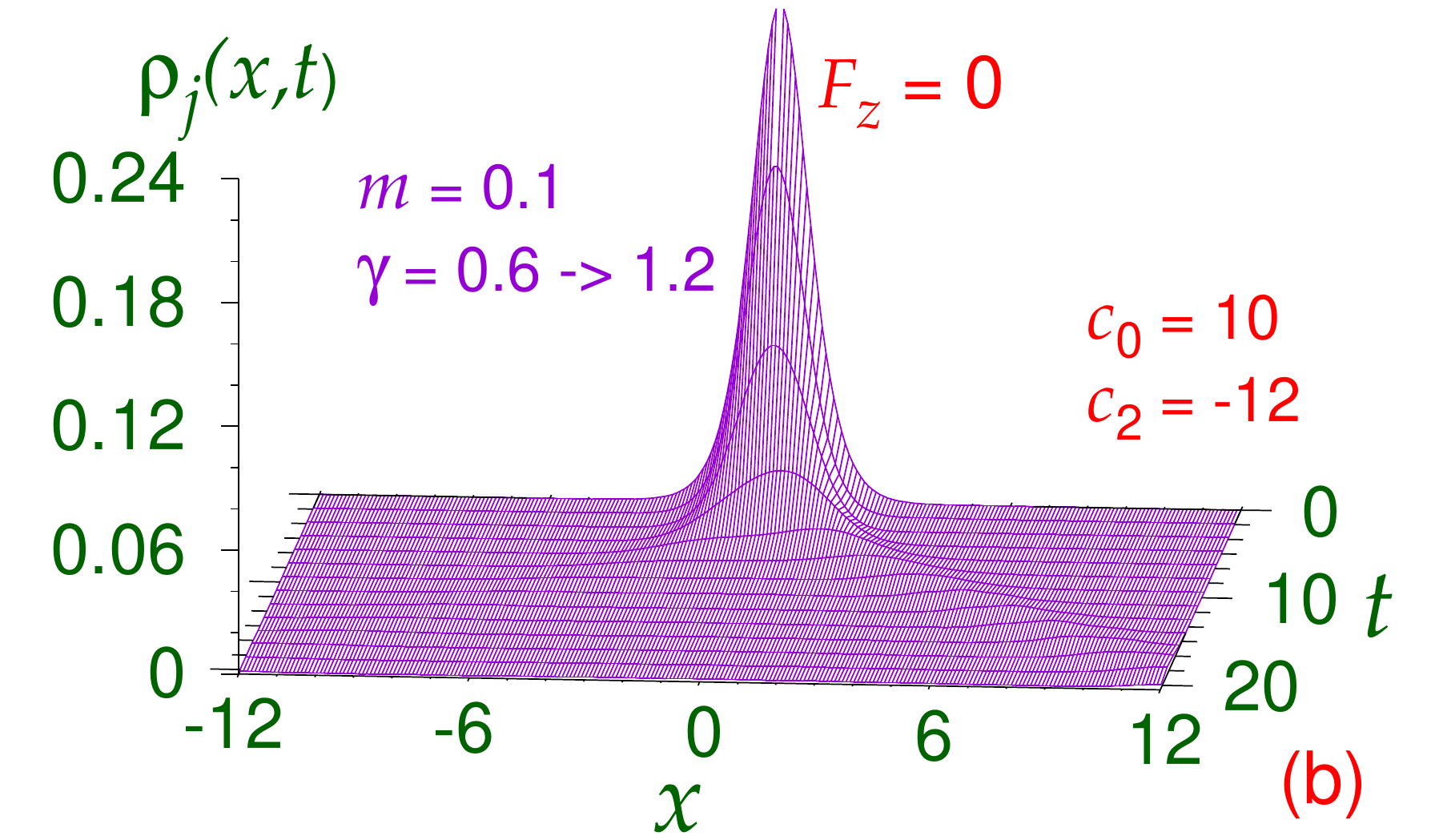}
 
\caption{(Color online) Dynamics of phase separation of a three-component vector soliton  upon changing the
 SO-coupling strength $\gamma$ from a value smaller than its critical value $\gamma _c$
$(=0.7071)$  for phase separation to a value larger than 
$\gamma_c$ by plots of densities of components (a) $F_z=\pm 1$ and (b)  $F_z=0$ versus $x$ and $t$, obtained by real-time propagation. Employed parameters are 
$ c_0=10, c_2=-12, m=0.1, \gamma_c = 0.7071$, while $\gamma $ was changed from 0.6  to 1.2.   }
\label{fig8} \end{center}
\end{figure}

Next we demonstrate the dynamical phase separation of a  three-component vector soliton. For this purpose we consider the ground-state three-component vector soliton profile
soliton profile for   $c_0=10, c_2=-12, m=0.1, \gamma =0.6$ obtained by imaginary-time propagation  and  subject it  to 
real-time propagation upon changing $\gamma$ to 1.2.  The initial $\gamma< \gamma_c= 0.7071 $  is appropriate for the formation of 
an overlapping three-component vector soliton in the ground state and the final $\gamma$ is appropriate for a phase-separated binary vector soliton in the ground state. In Figs. \ref{fig8}(a) and (b) we display the time evolution of density 
of the components $F_z=\pm 1$ and $F_z=0$, respectively. The components $F_z=\pm 1$ move away from each other while the component $F_z=0$ vanishes after time evolution as displayed in Figs. \ref{fig8}(a)-(b).

 \subsection{Collision of moving solitons}

\begin{figure}[!t]
\begin{center}
\includegraphics[trim = 0mm 0mm 0cm 0mm, clip,height=.5\linewidth,width=\linewidth, clip]{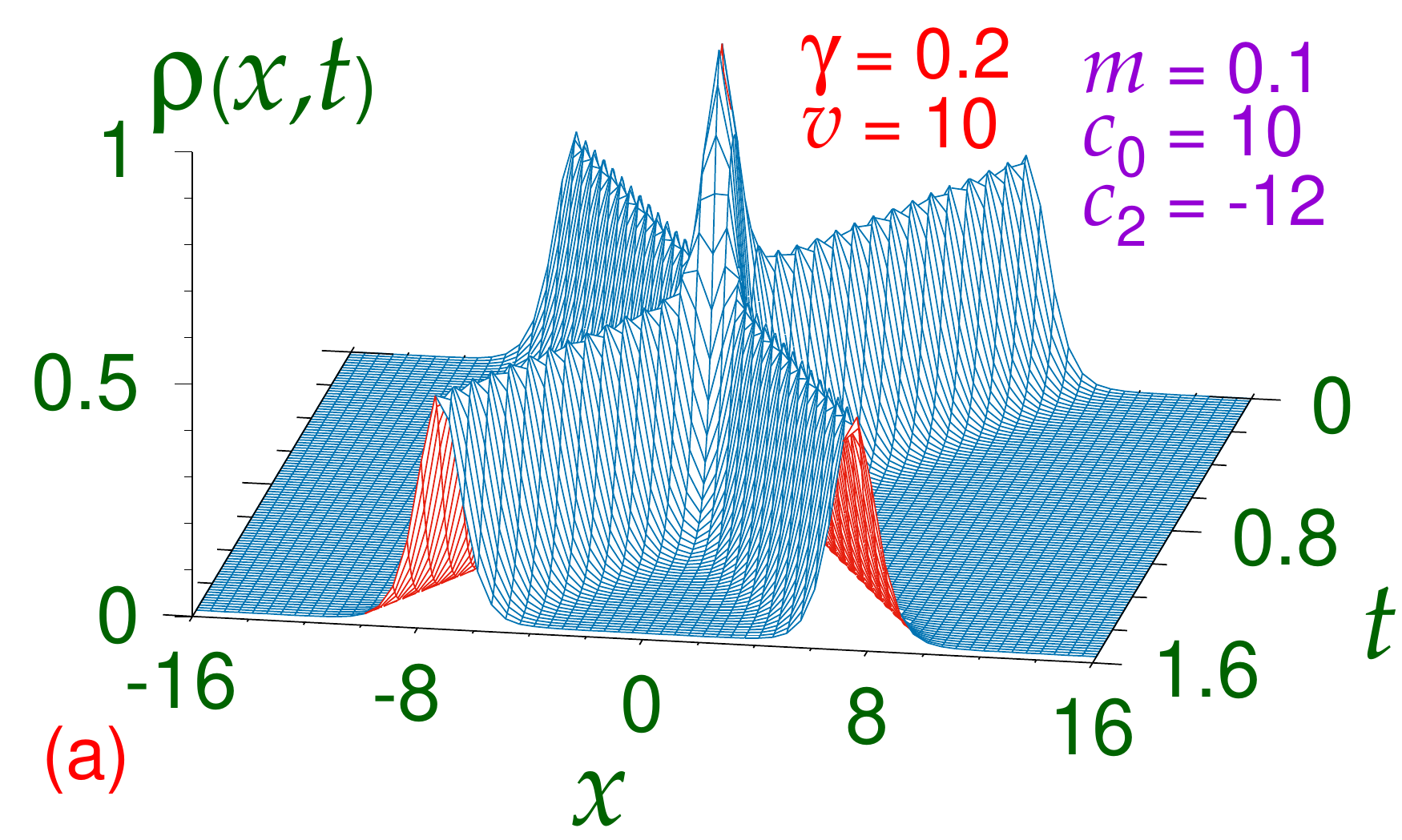}
\includegraphics[trim = 0mm 0mm 0cm 0mm, clip,height=.5\linewidth,width=\linewidth,clip]{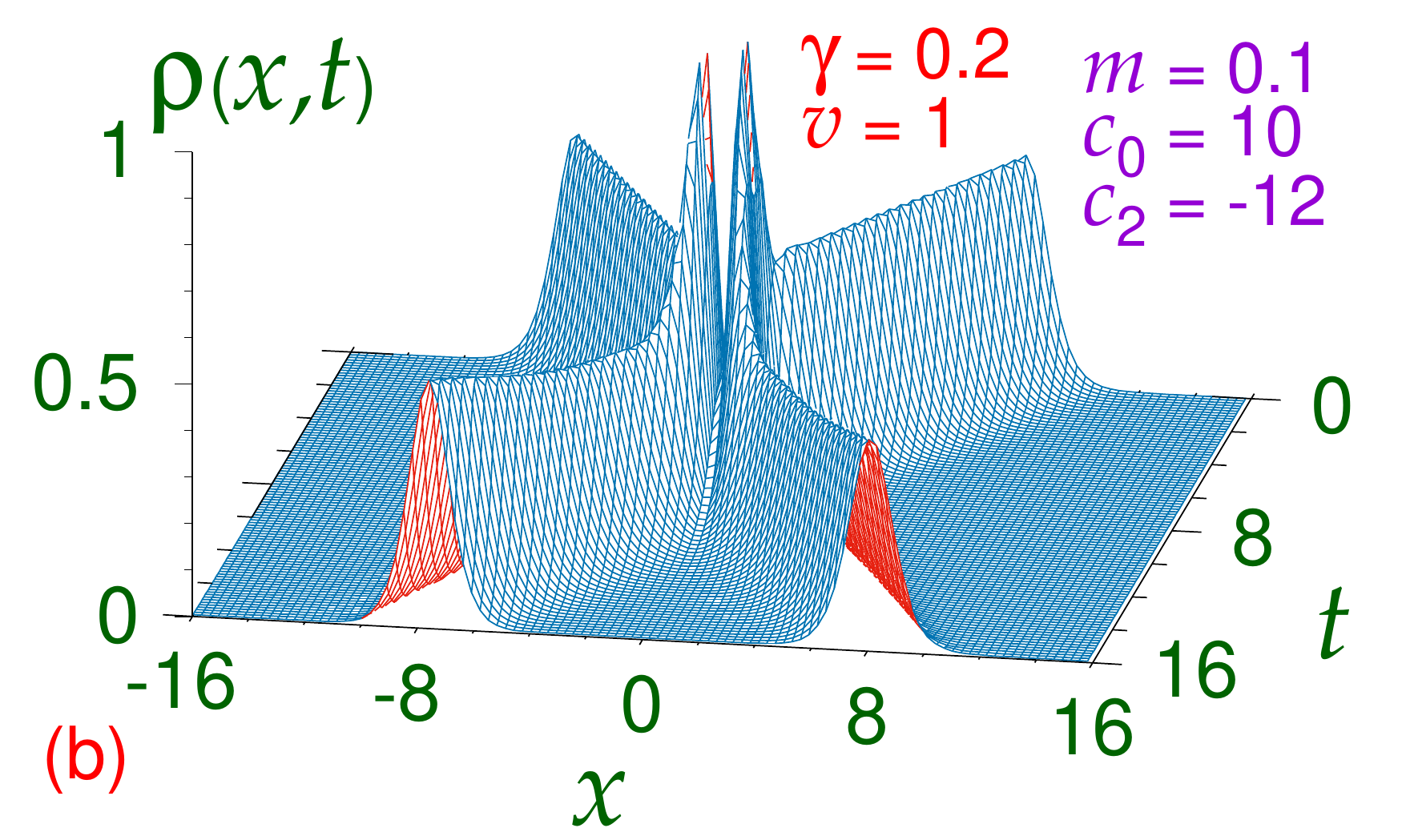}
 \includegraphics[trim = 0mm 0mm 0cm 0mm, clip,height=.5\linewidth,width=\linewidth,clip]{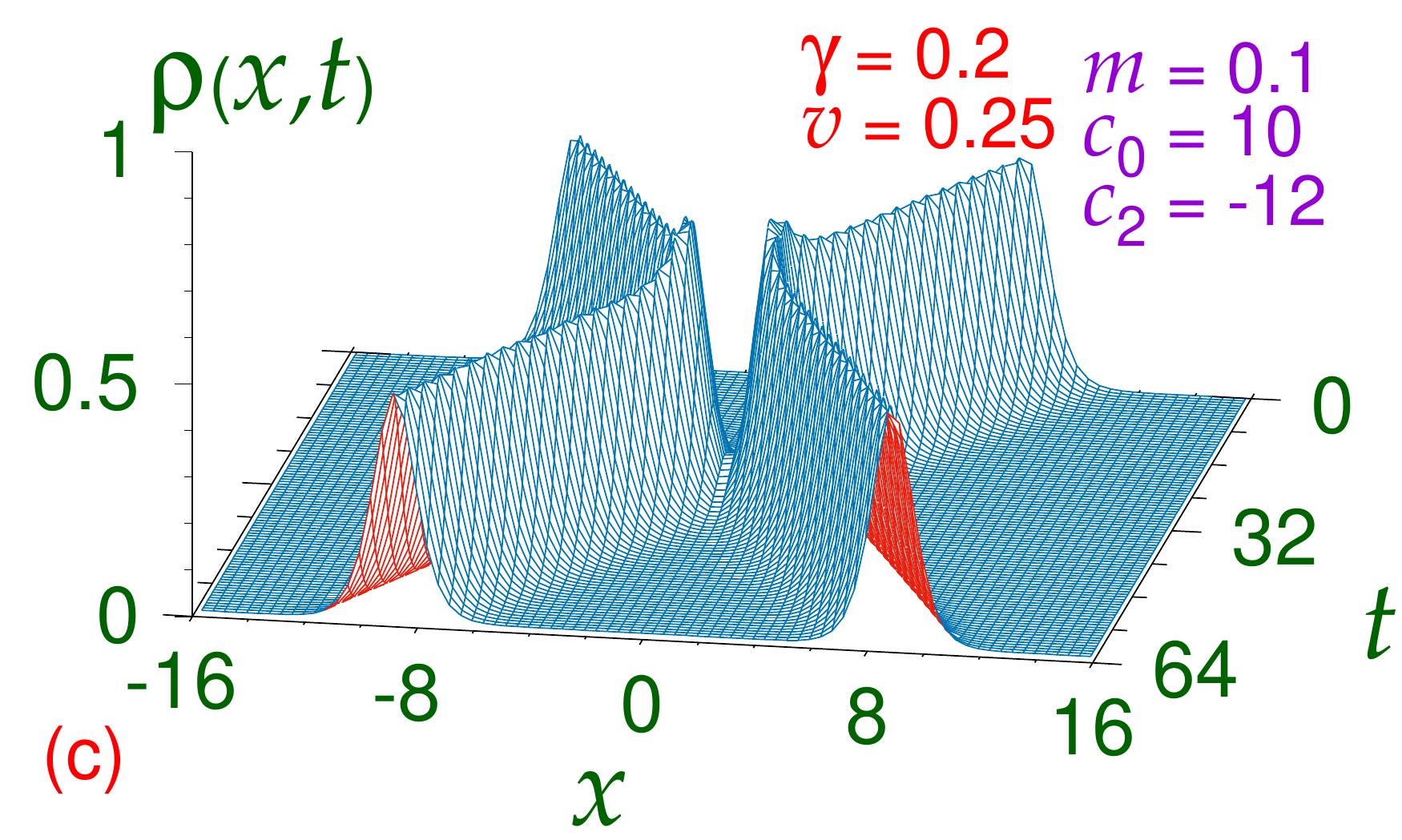}
 \includegraphics[trim = 0mm 0mm 0cm 0mm, clip,height=.5\linewidth,width=\linewidth,clip]{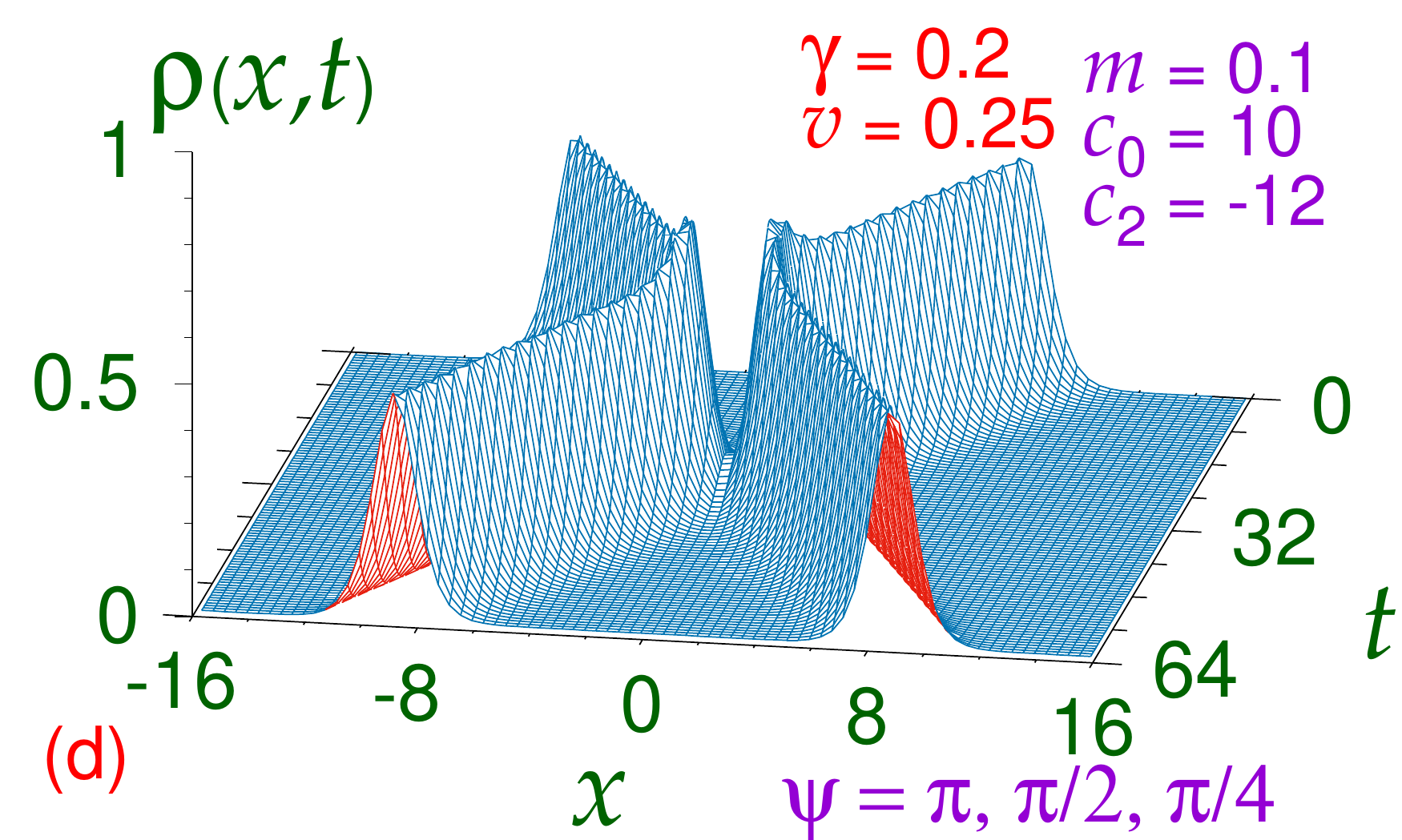}
\caption{(Color online) (a) Elastic nature of collision  dynamics of two  vector  solitons of Fig. \ref{fig7}  with $\gamma=0.2$
illustrated through a plot of total density $\rho(x,t)$, obtained by real-time propagation, 
versus $x$ and $t$. At $t=0$ individual solitons are placed at $x=\pm 8$ and set into motion 
in opposite directions with velocity $v=\pm 10$ by multiplying  the imaginary-time wave functions 
of the two vector solitons by $\exp(\pm ivx)$, respectively.   The same dynamics with (b) $v=\pm 1$
and (c)  $v=\pm 0.25$. (d) The dynamics of (c) after introducing a relative phase $\psi=\pi, \pi/2, \pi/3$ or $\pi/4$ between the two vector solitons.   Other parameters are $c_0=10, c_2=-12, \gamma =0.2, m=0.1$.}
\label{fig9} \end{center}
\end{figure}

\begin{figure}[!t]
\begin{center}
\includegraphics[trim = 0mm 0mm 0cm 0mm, clip,width=.9\linewidth,clip]{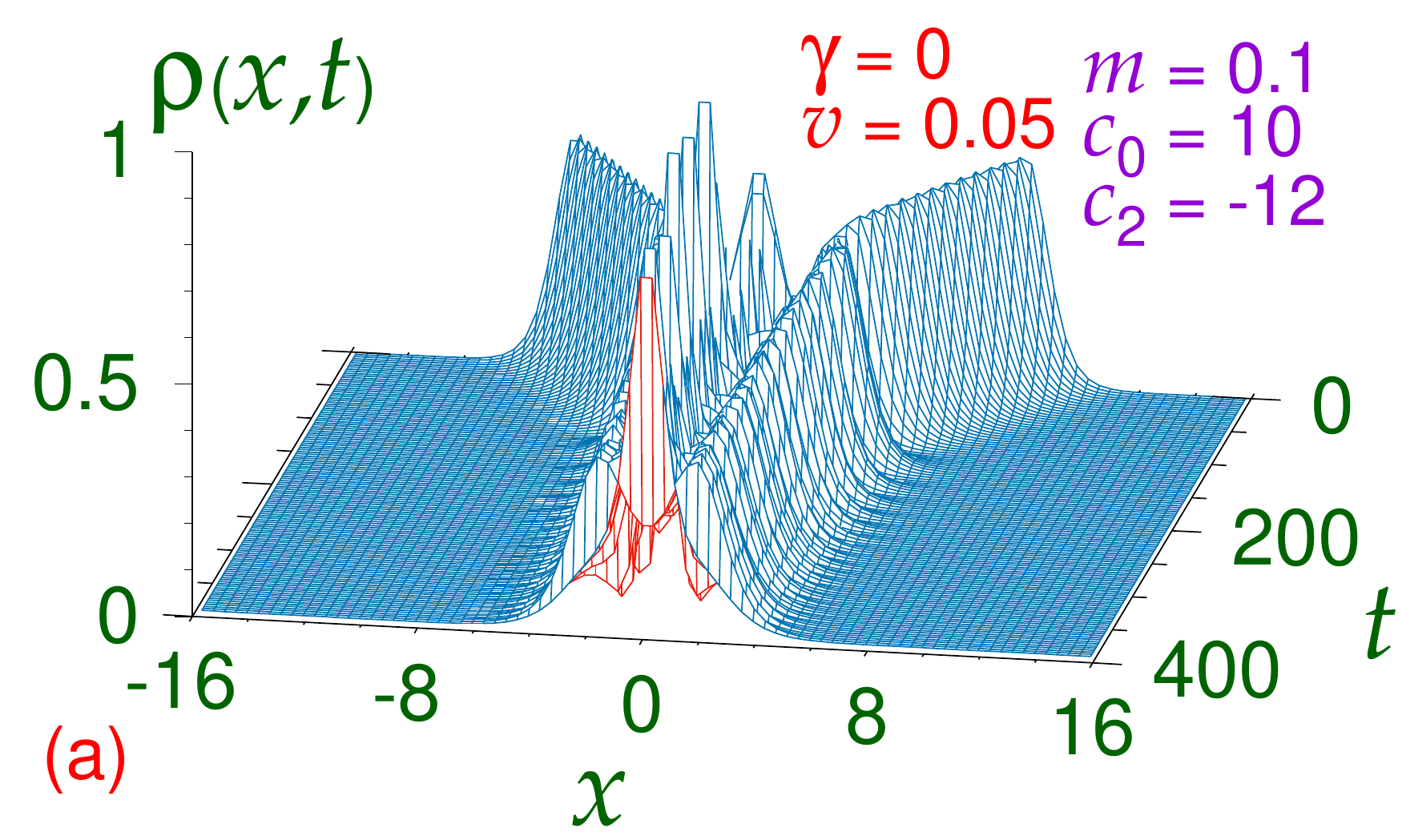}
\includegraphics[trim = 0mm 0mm 0cm 0mm, clip,width=.9\linewidth,clip]{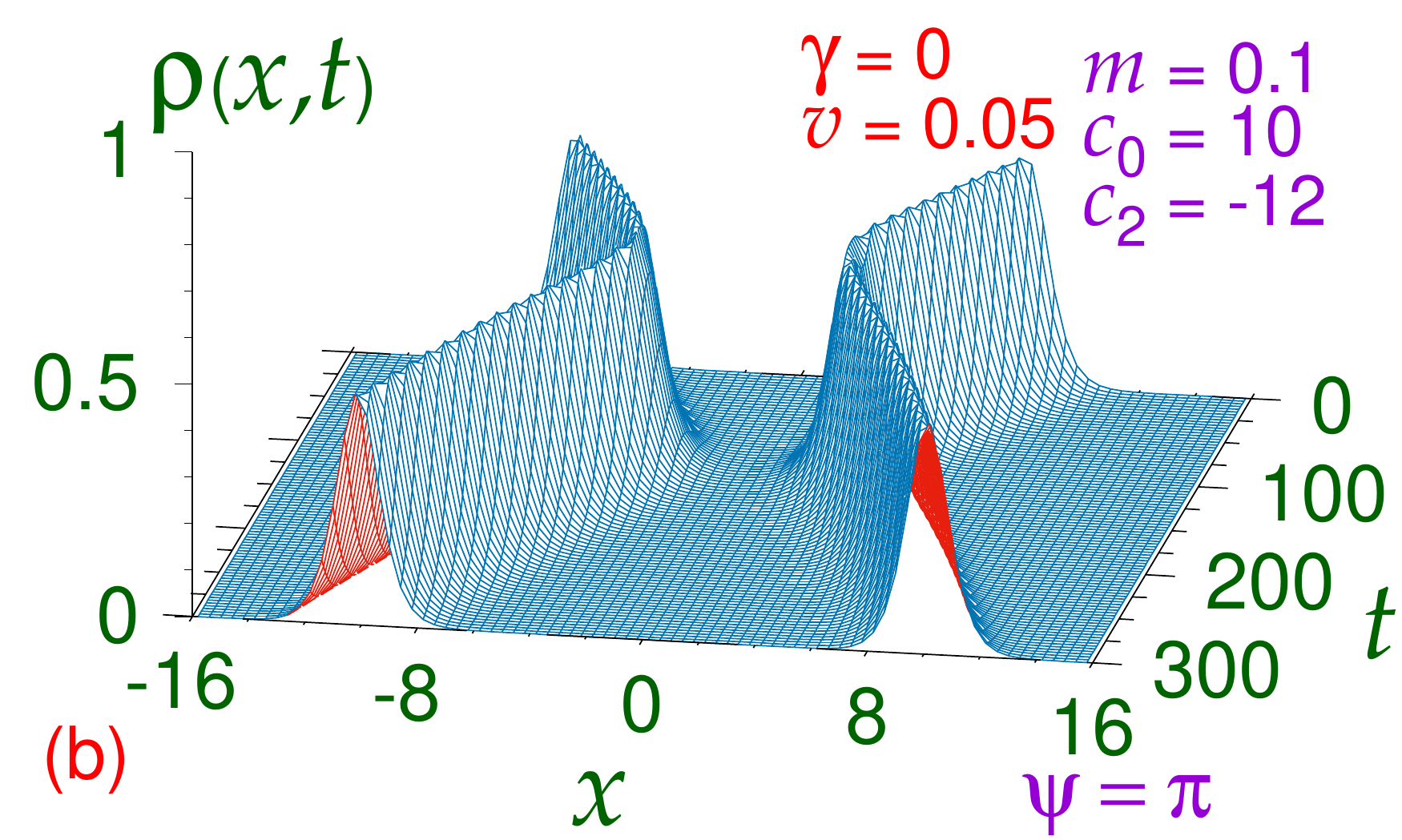} 
\includegraphics[trim = 0mm 0mm 0cm 0mm, clip,width=.9\linewidth,clip]{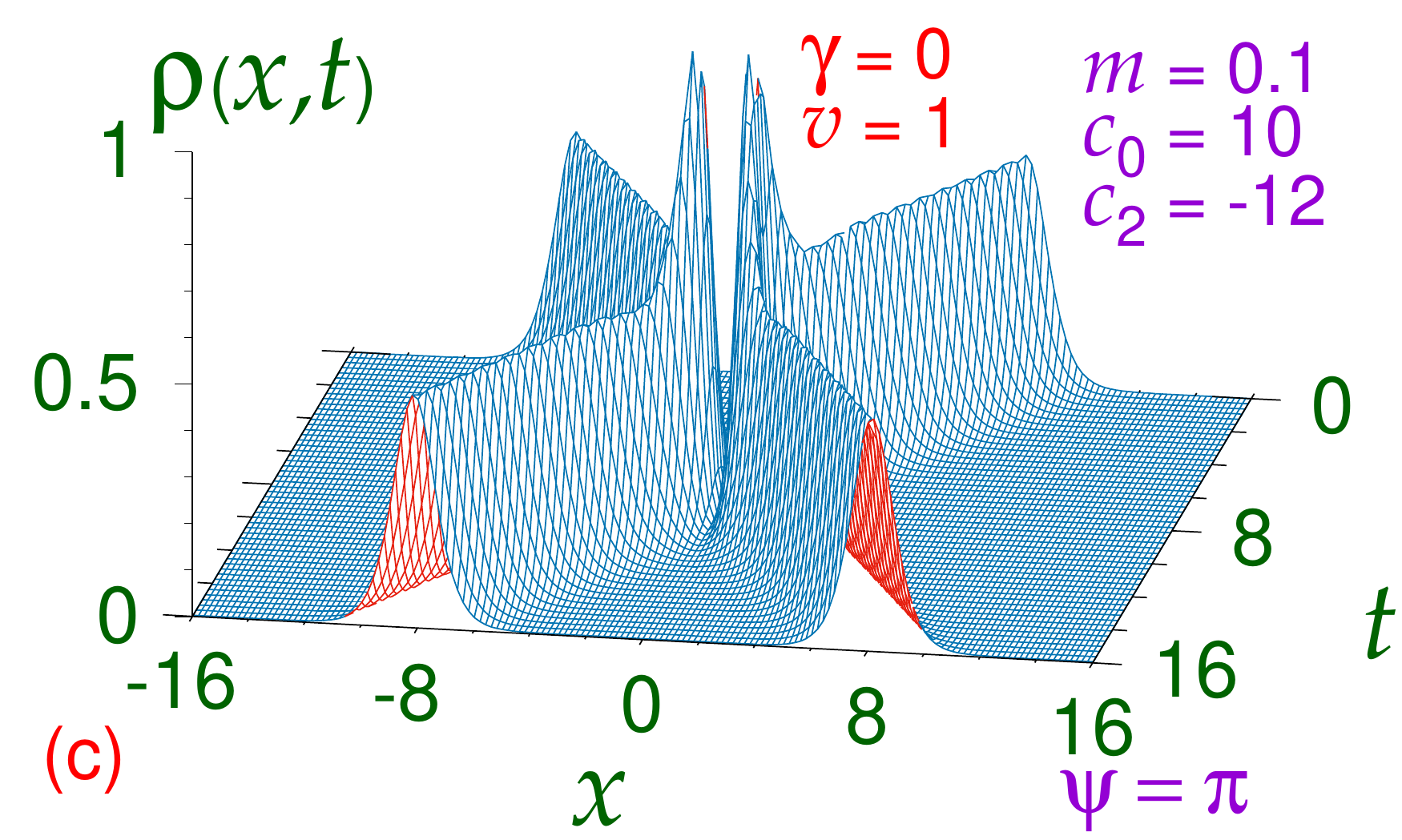} 
{
\caption{(Color online) (a) Inelastic  collision  dynamics resulting in soliton-molecule 
formation
of two  vector  solitons of Fig. \ref{fig7}  with SO coupling switched off ($\gamma=0$)
illustrated through a plot of total density $\rho(x,t)$, obtained by real-time propagation, 
versus $x$ and $t$. At $t=0$ individual solitons are placed at $x=\pm 8$ and set into motion 
in opposite directions with velocity $v=\pm 0.05$. (b)  The  dynamics of (a)  with   a relative phase $\psi +\pi$ introduced between the two vector solitons at $t=0$. 
(c) The  dynamics of (a)  with  a velocity $v=1$ and  a relative phase  $\psi=\pi$  between the two vector solitons at $t=0$. 
Other parameters are $c_0=10, c_2=-12,
 \gamma=0,  m=0.1$.}

}
\label{fix} \end{center}
\end{figure}

The dynamics of moving solitons  is next studied by  first generating
the ground-state overlapping vector  soliton numerically using imaginary-time propagation. 
The complex wave-function components, so obtained, are then multiplied by a complex phase $\exp(ixv)$,
which are used as the initial states in real time simulation. The generated vector soliton is 
 a moving soliton with velocity $v$ in the limit of very small space and time steps $dx$ and $dt$.  In the following calculation
the employed values of space and time steps 
  are $dx=0.025$ and  $dt=10^{-5}$, respectively.  {We will demonstrate that 
the present vector  soliton can move without changing the density profile of
the (three) components.} The collision dynamics  of  scalar
condensates depends \cite{referee}  on the relative velocity, relative amplitude, and relative phase $\psi$ of the two colliding solitons.  In   the present collision of SO-coupled  vector solitons we find that the relative velocity   has considerable effect on the collision process. However, the collision is  reasonably insensitive to the relative phase $\psi$ and  relative amplitude of the soliton  components. 
 To demonstrate the solitonic property of the overlapping vector soliton, we 
study the collision of two vector solitons each generated by imaginary-time propagation with parameters
$c_0=10, c_2=-12, m=0.1, \gamma=0.2$.   We take two vector solitons and place them at positions $x\equiv d=\pm 8$ 
and set them in motion in opposite directions    with velocity $v=\pm 10$ {(relative velocity of 20)}  so as to collide  at 
$x=0$ after time $t=d/v=0.8$. { During  collision,   the solitons are found to pass through each other essentially unchanged, and we 
study the collision dynamics. Both  magnetization and normalization are conserved during  
propagation of a  vector soliton resulting in the conservation of density profiles of each component.
The}  density of each vector soliton is conserved after collision showing
its elastic nature.  This is displayed in Fig. \ref{fig9}(a) via a plot of total density $\rho(x,t) $ 
 of the two vector solitons during collision. The two solitons emerge with the same velocity  and the same total 
density after collision.  { However, the SO coupling generates a repulsion between the solitons and the situation changes 
at small velocities $v$ as illustrated in  Figs. \ref{fig9}(b) and (c)  at  velocities $v=1$ and 0.25, respectively. The results of collision confirm a repulsion between the two SO-coupled  vector solitons at small velocities and at small separation between them.
For $v=0.25$ in Fig. \ref{fig9}(c), the repulsion due to SO-coupled repulsive interaction  stops the two vector solitons from meeting each other; they come close at $t=d/v= 32$,  turn back and move away from each other with the same speed. 
 This repulsion is also indicated for  $v=1$ in Fig. \ref{fig9}(b), where the two vector solitons show a tendency to stay apart and not mix with each other.  At all velocities, the two SO-coupled vector solitons emerge after collision without any visible deformation demonstrating the  elastic nature of collision at all velocities.

{
To study the effect of relative phase  of the two colliding vector solitons, we repeated the collision dynamics exhibited in Figs. \ref{fig9}(a)-(c) with  relative phases of $\pi, \pi/2, \pi/3$ and $\pi/4$ between the two colliding vector solitons. The dynamics of     Figs. \ref{fig9}(a)-(c) remains unchanged after the introduction of the relative phase  demonstrating no effect of phase on the collision. 
In Fig. \ref{fig9}(d) we plot the collision dynamics of Fig. \ref{fig9}(c) after introducing the 
relative phase $\psi=\pi$ between the components at $t=0$. From Figs. \ref{fig9}(c)-(d) we find that the 
relative phase has no effect on the collision. The phases $\psi = \pi/2, \pi/3$ and $\pi/4$ were also
found to yield the same dynamics. 
In case of usual scalar solitons,  the effect of relative phase on collision dynamics is the maximum    
for  $\psi=\pi$. The attractive interaction between two scalar solitons become repulsive upon the introduction of a relative phase of $\pi$ between the two colliding solitons \cite{referee}.  
}

To be sure that the small SO-coupled repulsive interaction $(\gamma=0.2)$ is causing the solitons to turn back 
and move away in Fig. \ref{fig9}(c), we studied the collision dynamics  of vector solitons with the SO coupling interaction switched off. 
The resultant collision dynamics for a small velocity $v=0.05$ and SO coupling $\gamma = 0$ is displayed in Fig. 10(a).  In this case the 
vector solitons come close, meet each other at time $t=d/v=160$, and form a vector soliton molecule in an excited state which
executes breathing oscillation and  never separate showing the presence of an attraction between the two vector solitons.  To study the effect of relative phase between the two vector solitons without SO coupling, we repeated the collision dynamics of Fig. 10(a) with a relative phase of $\pi$ between the two vector solitons and for $\gamma =0$ and the result is illustrated in Fig. 10(b). We consider a relative phase of $\pi$, as for this phase separation the effect of relative phase on collision dynamics 
is expected to be largest. 
 In the phase-changed configuration there is a strong repulsion between the two vector solitons which keep them apart.
{As in the case of scalar solitons,  the two vector solitons come close and turn back and emerge with the same speed without deformation as shown in 
   Fig. 10(b). To see the  effect of relative phase on velocity $v$, we consider in Fig. 10(c) the collision of the two solitons of Fig. 10(b)   with an initial velocity $v=1$ and a relative phase $\psi=\pi$.
 Even at a larger velocity, when the effect of relative phase is expected to be less,  the two vector solitons avoid each other because of the repulsion introduced by the relative phase.
 }

}

\section{Summary}
\label{Sec-IV}

We studied the generation, phase separation, and collision dynamics   of overlapping three-component  $(F_z=0,\pm 1)$ quasi-1D 
 vector solitons of a
SO-coupled  spin-1
spinor ferromagnetic BEC ($c_2< 0$) by a numerical solution and 
an analytic approximation of the mean-field GP equation, the SO coupling being of the form $\gamma p_x \Sigma_z$.   The solitons appear for net interaction strength   $(c_0+c_2)<0$.  
  The phase separation of the $F_z=\pm 1$ components  takes place for the strength $\gamma$ of  SO coupling  $p_x\Sigma_z$ above a 
critical value $(\gamma> \gamma_c)$, while the $F_z=0$  component vanishes.  
The vector solitons are demonstrated to be mobile and stable. 
By real-time simulation, we demonstrated the dynamical phase separation of an overlapping vector soliton upon increasing 
the strength of SO coupling above the critical value, viz. Fig. \ref{fig8}. At all velocities, 
the collision dynamics between two such vector solitons is found to be elastic with the conservation of the 
 densities of each individual vector solitons, viz. Figs. \ref{fig9}. {   The vector solitons repel each other due to SO coupling  and consequently,  in collision at  small velocities, the vector solitons come close to reach other and bounce back with same speed  without ever meeting each other as shown in Fig. \ref{fig9}(c). The collision dynamics of two SO-coupled vector solitons is found to be insensitive to the relative phase between them.
In the absence of SO coupling,  in collision at small velocity the two vector solitons attract each other and form a vector soliton molecule in an excited state and never separate, viz. Fig. 10(a), quite similar to collision of two multi-component scalar solitons. Also, in the absence of SO coupling the collision dynamics is very sensitive to the relative phase, viz. Fig. 10(b).   
With the present experimental know-how,  these vector solitons can be generated in a laboratory in a routine fashion 
and our predictions can be verified.


\begin{acknowledgements}
This work is financed by the Funda\c c\~ao de Amparo \`a Pesquisa do Estado de 
S\~ao Paulo (Brazil) under Contract Nos. 2013/07213-0, 2012/00451-0 and also by 
the Conselho Nacional de Desenvolvimento Cient\'ifico e Tecnol\'ogico (Brazil).
\end{acknowledgements}

\end{document}